\documentclass[acmlarge, nonacm]{acmart}
\usepackage{multirow}
\usepackage{tabularx}
\AtBeginDocument{%
  }

\setcopyright{acmlicensed}
\copyrightyear{2025}
\acmYear{2025}
\acmDOI{XXXXXXX.XXXXXXX}
\acmConference[CSCW '26]{October
  2026}{North America}

\newcommand{\highlight}[1]{#1}

\begin{document}

\title[Where's the Team Spirit?]{Where's the Team Spirit? An Exploratory Study on Team Development Through Co-located Tablet-Based VR}



\author{Irina Paraschivoiu}
\authornote{Corresponding author.}
\email{irina.paraschivoiu@polycular.com}
\affiliation{%
  \institution{Paris Lodron University of Salzburg}
  \city{Salzburg}
  \country{Austria}
}
\affiliation{%
  \institution{Polycular}
  \city{Salzburg}
  \country{Austria}
}

\author{Thomas Layer-Wagner}
\email{thomas.layer-wagner@polycular.com}
\affiliation{%
  \institution{Polycular}
  \city{Salzburg}
  \country{Austria}
}

\author{Klaus Neundlinger}
\email{klaus.neundlinger@in-scope.com}
\affiliation{%
  \institution{In Scope}
  \city{Salzburg}
  \country{Austria}
}

\author{Simone Rack}
\email{simone.rack@in-scope.com}
\affiliation{%
  \institution{In Scope}
  \city{Vienna}
  \country{Austria}
}

\author{Markus Tatzgern}
\email{markus.tatzgern@fh-salzburg.ac.at}
\affiliation{%
  \institution{Salzburg University of Applied Sciences}
  \city{Salzburg}
  \country{Austria}
}

\renewcommand{\shortauthors}{Paraschivoiu et al.}

\begin{abstract}
We explore how narrative-driven asymmetric VR experiences can support the development of teamwork-related knowledge, skills, and attitudes (KSAs), such as communication, coordination, trust, and reflexivity. We present the design and evaluation of a tablet-based VR training experience structured around spatial separation, tool asymmetry, and interdependent tasks that require verbal coordination. The experience was designed based on interviews with HR professionals and mapped to a framework of established KSAs. We conducted a co-located user study (N=16) that involved two consecutive collaborative scenarios. Our findings show that users adapted dynamically using verbal exchange, role negotiation, and shared representations to coordinate under asymmetric conditions. We also observed active application of teamwork KSAs. Based on our insights, we present design recommendations for creating effective immersive team training interventions.
\end{abstract}



\received{May 2025, accepted April 2026 to ACM CSCW}


\begin{teaserfigure}
 \includegraphics[width=\textwidth]{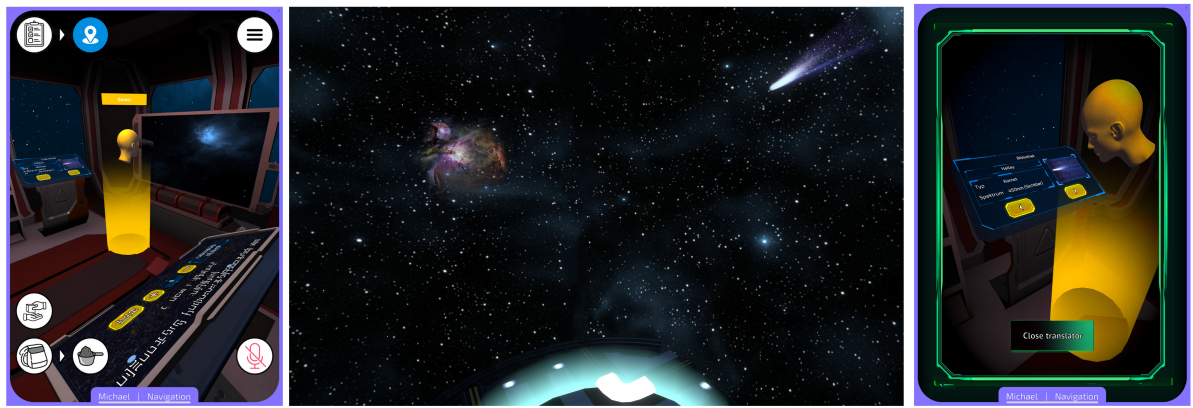}
  \caption{Illustrative collaboration scenario: one player retrieves distances to constellations (left and middle) based on the description provided by the other player (right).}
  \Description{The figure shows three images. The left image shows a player in the Space View, wearing the Infrared glasses. Center image: visual from inside the Space View, showing constellations and stars. Right image: player is looking at the display showing constellations from the ship's database}
  \label{Teaserfigure}
\end{teaserfigure}

\maketitle

\section{Introduction}
Organizations rely on teams to perform increasingly complex and distributed tasks. To succeed, teams must develop key skills, such as communication, coordination, and collective problem-solving, which can be trained through targeted interventions \cite{stevens_knowledge_1994, shuffler_theres_2011}. We refer to this process as \textit{team development}, the ongoing progression through which teams improve their collaborative performance.

To describe this progression, Tuckman conceptualized a model consisting of five stages: \textit{forming}, \textit{storming}, \textit{norming}, \textit{performing}, and \textit{adjourning} \cite{tuckman_developmental_1965, tuckman_stages_1977}. In the initial stages (\textit{forming}), teams set goals and a general structure, including roles, responsibilities, and tasks. Usually, this is followed by a more volatile phase (\textit{storming}), where they renegotiate roles and may question leadership or procedures. Cohesion is generally formed during \textit{norming}, while in the \textit{performing} stage teams evolve to high-functioning collaboration. Although some aspects of teamwork performance are related to personal traits (e.g. extraversion \cite{west_effective_2006}), research has shown that \textit{knowledge, skills and abilities (KSAs)}, such as clarity, coordination, and communication, can be improved through targeted interventions \cite{stevens_knowledge_1994, shuffler_theres_2011, lacerenza_team_2018, shuffler_developing_2018, sottilare_designing_2018}.  

Two common types of such  interventions are \textit{team training}, which targets performance-relevant skills (e.g. decision-making, coordination) and \textit{team building}, which fosters relational attitudes such as trust and cohesion \cite{shuffler_theres_2011}. When carefully designed, both can yield measurable improvements in team effectiveness: \textit{"When delivering training programs, it is critical that salient theory and evidence are referenced to provide the most effective training"} \cite[p.521]{lacerenza_team_2018}.

Group discussions, simulation exercises, e-learning modules, and evaluation metrics are some of the many approaches to team training and team building. More recently, there has been growing interest in the potential of immersive game-based experiences for team training \cite{cherbonnier_collaborative_2024}. Such playful learning enables hands-on experiential engagement, placing participants in metaphorical or narrative-driven situations that mirror real-world collaboration. When these playful experiences are delivered through Virtual Reality (VR), they additionally benefit from its unique affordances: \textbf{immersion and presence} \cite{tran_survey_2024}, which create realistic social dynamics and emotional investment \cite{potts_retrosketch_2025}; \textbf{ecological validity} \cite{parsons_virtual_2015}, by simulating coordination under stress, time constraints, or ambiguity; \textbf{immediate feedback}, enabling “learning by doing” \cite{kolb_experiential_2015}; and \textbf{desirable difficulties} \cite{healy_material_2005}, such as restricted communication or asymmetric information, which encourage deliberate skill use. 

Although experiencing VR via a head-mounted display (HMD) is associated with increased levels of immersion and presence, HMDs remain rare in most organizational contexts. In contrast, handheld devices such as smartphones and tablets make screen-based VR widely accessible, cost-effective, and easier to deploy at scale \cite{ouverson_composite_2021}. This makes it a promising platform for immersive team training interventions.

In this paper, we explore how a narrative-driven, tablet-based VR experience (Figure \ref{Teaserfigure}) can support the development of teamwork KSAs in co-located teams. We designed and evaluated two collaborative experiences, through a process that included interviews with HR professionals and review of teamwork literature, with a focus on KSAs. In proposing our design, we deliberately focused on the mechanics that create \textit{desirable difficulties} to emphasize the need for users to practice KSAs related to teamwork, namely: \textbf{ spatial separation} (to promote verbal coordination), \textbf{asymmetric tools and information} (to enforce interdependence), and \textbf{structured debriefing} (to support team reflection). \highlight{We focus on co-located teams, as collaboration tends to be more effective when users share the same physical space, whereas remote settings introduce additional challenges related to awareness \cite{radu_survey_2021}.}

We make the following contributions:
\begin{itemize}
    \item We present results of our design process grounded in literature and interviews with HR professionals, leading to a set of features and mechanics aligned with teamwork KSAs.
    \item We present empirical findings from a co-located study with 16 participants, triangulating questionnaire data, video analysis, and group reflections to understand the collaboration of participants when using the experience.
    \item Based on our findings, we derive design recommendations to develop immersive, tablet-based VR experiences that support team development, including guidelines on (1) communication design, (2) interdependence, (3) role metaphors, and (4) structured reflection. 
\end{itemize}

\section{Related Work}

\subsection{Team Development}
Teamwork has long been a central focus in the social sciences \cite{tuckman_stages_1977, west_effective_2006}. Organizations rely on teams to achieve their goals, and the effectiveness of team collaboration has been shown to impact performance in complex tasks \cite{shuffler_developing_2018} and overall productivity \cite{gallup_inc_state_2024}. As a result, research has been concerned with understanding what kind of skills and knowledge individuals need to perform best in teams. Stevens and Campion \cite{stevens_knowledge_1994} proposed five essential knowledge, skills and abilities (KSAs) subcategories: \textbf{ conflict resolution, collaborative problem-solving, communication, goal setting}, and \textbf{task coordination}. These are further detailed in 14 specific KSAs that can be trained in workplace settings, for example, listening non-evaluatively or coordinating interdependent tasks \cite[p.505]{stevens_knowledge_1994}. Sottilare expanded on this by including constructs related to interpersonal relationship management, such as \textbf{collective efficacy, trust, and cohesion}, supported by behavioral markers (e.g. use of honest language or emotional expression) \cite{sottilare_designing_2018}.

A closely related concept to performance monitoring \cite{stevens_knowledge_1994} is \textbf{team reflexivity}: the ability of teams to reflect and adapt their strategies and behaviors \cite[p.153]{konradt_reflexivity_2016}. Reflexivity typically occurs during the transition phases, preparing teams for future work \cite[p.162]{konradt_reflexivity_2016} and is closely related to cycles of adaptation and learning \cite{burke_understanding_2006}. This process is essential to improve shared mental models and situational awareness \cite{konradt_reflexivity_2016, burke_understanding_2006}. Table \ref{tab:ksa_categories}  presents a consolidated list of KSAs for teamwork based on Stevens \cite{stevens_knowledge_1994}, Sottilare \cite{sottilare_designing_2018} and Konradt \cite{konradt_reflexivity_2016}. KSAs 1–14 are taken directly from Stevens and Campion, using their original numbering and structure. The authors extracted additional KSAs from the other sources and continued the sequence. The wording has been slightly edited for conciseness while preserving the original intent.

\begin{table}
\caption{Teamwork Knowledge, Skills, and Abilities (KSAs) adapted from literature}
\label{tab:ksa_categories}
\setlength{\extrarowheight}{.4em}
\begin{tabular}{p{7em}p{3em}p{28em}}
\toprule
\textbf{Category} & \textbf{ID} & \textbf{Description of KSAs} \\
\midrule

\multirow[t]{3}{=}{\textbf{Conflict Resolution} \cite{stevens_knowledge_1994}}
& KSA1 & Recognize and encourage desirable, but discourage undesirable conflict. \\
& KSA2 & Recognize the type and source of conflict confronting the team and implement an appropriate conflict resolution strategy. \\
& KSA3 & Employ an integrative (win-win) negotiation strategy. \\

\multirow[t]{2}{=}{\textbf{Collaborative Problem-Solving} \cite{stevens_knowledge_1994}}
& KSA4 & Identify situations requiring participative group problem-solving and utilize the proper degree and type of participation. \\
& KSA5 & Recognize the obstacles to collaborative group problem-solving and implement appropriate corrective actions. \\

\multirow[t]{5}{=}{\textbf{Communication} \cite{stevens_knowledge_1994}}
& KSA6 & Understand communication networks and utilize decentralized networks to enhance communication. \\
& KSA7 & Communicate openly and supportively. \\
& KSA8 & Listen nonevaluatively and use active listening techniques. \\
& KSA9 & Maximize consonance between nonverbal and verbal messages. \\
& KSA10 & Engage in small talk and recognize its importance. \\

\multirow[t]{4}{=}{\textbf{Goal Setting} \cite{stevens_knowledge_1994}}
& KSA11 & Establish specific, challenging, and accepted team goals. \\
& KSA12 & Monitor, evaluate, and provide feedback on both overall team performance and individual team member performance. \\
& KSA13 & Coordinate and synchronize activities, information, and task interdependencies between team members. \\
& KSA14 & Establish task and role expectations and ensure balancing of workload. \\

\multirow[t]{5}{=}{\textbf{Collective Efficacy} \cite{sottilare_designing_2018}}
& KSA15 & Reinforce the team’s ability to complete tasks successfully. \\
& KSA16 & Acknowledge contributions from teammates and integrate their input. \\
& KSA17 & Actively request and offer backup when needed. \\
& KSA18 & Express confidence and willingness to resolve disagreements. \\
& KSA19 & Belief in the team’s ability to accomplish goals. \\

\multirow[t]{4}{=}{\textbf{Trust} \cite{sottilare_designing_2018}}
& KSA20 & Share information and delegate tasks without excessive oversight. \\
& KSA21 & Seek input from others and accept feedback constructively. \\
& KSA22 & Avoid withholding task-related information. \\
& KSA23 & Offer help and accept backup behaviors as signs of mutual trust. \\

\multirow[t]{3}{=}{\textbf{Cohesion} \cite{sottilare_designing_2018}}
& KSA24 & Participate actively in shared team rituals and affirm team identity. \\
& KSA25 & Promote inclusive participation and value every team member’s input. \\
& KSA26 & Support the social fabric of the team by encouraging informal interaction and camaraderie. \\

\multirow[t]{2}{=}{\textbf{Reflexivity} \cite{konradt_reflexivity_2016}}
& KSA27 & Generate alternative strategies and compare their potential impacts. \\
& KSA28 & Adjust strategies in response to performance or changes. \\

\bottomrule
\end{tabular}
\end{table}
\subsection{Asymmetric Playful Experiences}
Multiplayer asymmetric experiences offer a compelling design space for collaboration. Asymmetry introduces structural interdependence, requiring players to collaborate meaningfully by compensating for each other’s limitations. Harris et al. \cite{harris_leveraging_2016} introduced a foundational framework that categorizes asymmetry based on the dimensions of \textit{mechanics, interdependence}, and \textit{aesthetics}. For example, in terms of mechanics, Harris suggests that experiences can leverage asymmetry of ability, challenge, interface, information, investment, and goal/responsibility. Moreover, combining these asymmetries with well-structured dependencies led to deeper cooperation, dynamic role negotiation, and frequent verbal coordination \cite{harris_leveraging_2016}. Users also reported that the need to depend on each other improved the enjoyment and quality of teamwork. 

Rogers et al. \cite{rogers_best-fit_2021} extended this framework in their systematic review of 25 asymmetric VR games, highlighting that information, interface, and responsibility asymmetries were most used to induce cooperation. They observed that bidirectional dependence was present in most collaborative games, promoting social connection and requiring constant communication to complete tasks. 

Asymmetric designs have been used to increase intergenerational connections between family members \cite{pais_promoting_2024} and friends \cite{depping_designing_2018}, as well as to create engaging experiences for users of diverse abilities \cite{goncalves_exploring_2021}. In families, asymmetry of information and interface encouraged discussion and discovery, but the lack of explicit feedback sometimes led to confusion and disengagement \cite{pais_promoting_2024}. For friends, asymmetry worked as the main catalyst for communication, and successful collaboration required players to trust the input and judgment of each other \cite{goncalves_exploring_2021}. However, some participants experienced tensions and fatigue due to repetitive communication patterns or unbalanced cognitive loads. 

Asymmetric setups have also focused on engaging non-HMD users with VR players, to break up the potential loneliness effects of VR. Gugenheimer et al. \cite{gugenheimer_facedisplay_2018} and Zhou et al.\cite{zhou_astaire_2019} explored asymmetric setups involving one VR player and one non-HMD player, revealing that asymmetry embedded in game design can create novel types of experiences. However, perceived power imbalance can hinder collaboration unless it is intentionally embedded in the narrative or interface. Their studies showed that embracing asymmetry can transform imbalance into an opportunity for meaningful role differentiation.

In these studies, asymmetric play was shown to increase perceived social presence compared to symmetric play \cite{harris_asymmetry_2019}, reduce player loneliness \cite{liszio_influence_2017}, result in positive measurements of empathy and behavioral involvement \cite{sajjadi_maze_2014}, and communication and collaboration \cite{zhang_dice_2025}. Experiences that also integrate physicality create a strong presence through physical play \cite{wang_solo_2025, zhou_astaire_2019}. When the interdependency of the roles was high, these effects were stronger, but the players also appreciated variations in the degree of coordination required, suggesting that fluctuating interdependence might sustain engagement and reduce fatigue \cite{harris_asymmetry_2019}.

Although these studies did not specifically aim at team training, many of their positive effects are strongly related to these constructs. In our research, we build on this previous work by connecting such mechanics with KSAs from the previously listed teamwork literature to determine which designs can support their development.  

\subsection{Coupling Styles in Co-located Collaboration}
A substantial body of work in CSCW has explored co-located collaboration \cite{ens_revisiting_2019, lukosch_collaboration_2015}, particularly around shared surfaces and interactive environments. Descriptive frameworks and models have been proposed to characterize collaboration in settings such as digital tabletops and shared displays \cite{isenberg_co-located_2012, scott_territoriality_2004, tang_collaborative_2006}, hybrid or cross-device systems \cite{plank_is_2017, brudy_investigating_2018, schroder_collaborating_2023, jetter_transitional_2021, pointecker_bridging_2022} and digital-physical workspaces \cite{mai_evaluating_2018}. Especially relevant to this work are contributions that have focused on handheld devices \cite{olsson_technologies_2020, paraschivoiu_crafting_2025, wells_collabar_2020, poretski_physicality_2021}. Models for hybrid environments \cite{neumayr_domino_2018} have expanded on these to include transitions in partially distributed teams.

Given our focus on developing teamwork KSAs, these models provide a valuable methodological foundation. In particular, they help us interpret how participants coordinate in immersive collaborative environments and what coupling styles reveal about teamwork efficiency and skill development.

In the CSCW literature, \textit{coupling} refers to the degree of interdependence between users during joint activity \cite{baker_empirical_nodate, tang_collaborative_2006}. \textit{Tightly coupled} collaboration is typically required when team members must support each other to progress (e.g. solving interdependent puzzles), whereas \textit{loosely coupled} work occurs when tasks can be efficiently split \cite{neumayr_domino_2018, poretski_physicality_2021, tang_collaborative_2006, baker_empirical_nodate}. Previous studies have shown that individuals switch between these modes fluidly based on task demands, available resources, and interface design \cite{tang_collaborative_2006, baker_empirical_nodate, poretski_physicality_2021}.

More detailed classifications of coupling have emerged from observational studies. For example,\textit{View Engaged} refers to one participant working while another observes; \textit{Different Problem} denotes loosely coupled work in which individuals pursue separate tasks \cite{neumayr_domino_2018}. For our study, we adopt the \textit{Domino framework} from Neumayr et al. \cite{neumayr_domino_2018}, which extends previous models by introducing a richer set of interaction categories: six tightly coupled and three loosely coupled styles. Crucially, \textit{Domino} also offers a method for analyzing subgroup formations within larger teams. 

In addition to modeling collaboration, CSCW has also proposed key design principles to support co-located group work, including maintaining awareness, coordinating attention and instructions, supporting object manipulation, sharing environments, and preserving privacy \cite{radu_survey_2021}. However, when designing a team training intervention, we argue that intentionally violating some of these principles may be productive. For example, reducing awareness or introducing asymmetries \cite{ouverson_composite_2021} can create \textbf{desirable difficulties} that prompt learners to practice communication, clarification, and active listening.

In the context of team training, coupling styles offer more than a descriptive account of collaboration. They provide observable proxies for how participants practice and apply KSAs related to teamwork. For example, tightly coupled behaviors can reflect active communication and mutual support. By analyzing shifts in coupling styles during use, it is possible to see how different design elements foster (or hinder) collaborative skill development in practice.

\subsection{Research gap and Research Questions}
This paper addresses the increasing demand for team training \cite {lacerenza_team_2018} by investigating how collaborative VR experiences can train key teamwork skills. We focus on knowledge, skills, and attitudes (KSAs) essential for effective team performance, such as communication, coordination, and reflexivity. Research questions (RQ):
\begin{itemize}
    \item \textbf{RQ1}: How should a narrative-driven tablet-based VR experience be designed to support team training interventions? 
    \item \textbf{RQ2}: How do users  \highlight{enact} teamwork-related KSAs during the experience? 
\end{itemize}
Together, these questions bridge the gap between design research, immersive technology, and team training theory, enabling us to contribute both a functional prototype and actionable recommendations for designing collaborative KSA-focused VR systems. \highlight{Specifically, we focus on how these KSAs are enacted during collaborative activity within the experience, rather than on assessing their long-term development.}

\highlight{\section{Designing a Tablet-based VR Experience for Team Training}}
\label{sec:designing_vr}
Insights from interviews with HR professionals helped identify organizational challenges and training needs. We map these themes to established teamwork KSAs and derive a set of design requirements and features. Finally, we describe the collaborative VR scenarios.

\subsection{Expert interviews}
To address \textbf{RQ1}, we conducted semi-structured interviews with 11 HR professionals (M = 7, F = 4). Experts held mid- to senior roles and represented a variety of industries, including innovation consulting firms, research institutions, and large corporations in retail, logistics, and aviation. The recruitment was carried out through the professional networks of the researchers and through innovation support organizations. Table \ref{Experts} provides an overview of the participants.

\begin{table*}
  \caption{Overview of expert roles and industries}
    \label{Experts}
  \begin{tabular}{cccc}
    \toprule
    Participant code & Gender & Role & Industry\\
    \midrule
         P1 & Male & Senior innovation lead & Research and Development \\ 
        P2 & Male & Founder and CEO & Innovation consultancy \\ 
        P3 & Male & Founder and entrepreneur & Digitalization \\ 
        P4 & Male & Manager People and Culture & Retail \\ 
        P5 & Male & Head of Global HR & Logistics and automation \\ 
        P6 & Male & Founder and managing director & Educational technologies\\ 
        P7 & Male & Founder and expert & Coaching and HR consulting \\ 
        P8 & Female & Team lead, AI/ML & Aviation \\ 
        P9 & Female & Senior researcher & Collaboration and diversity \\ 
        P10 & Female & Founder and expert & Executive coaching \\ 
        P11 & Female & Founder and coach & Organisational development \\ 
  \bottomrule
\end{tabular}
\end{table*}

\subsubsection{Procedure}
Each interview lasted 30 to 40 minutes and was conducted remotely via video conferencing. Three researchers independently conducted the interviews. The interview protocol was organized into three sections:
\begin{enumerate}
\item {Current challenges related to team development.}
\item{Existing programs for team training and team building.}
\item{Future directions to enhance and support team development in organizations.}
\end{enumerate}

Participants were encouraged to speak from their personal experience. All participants provided their informed consent to participate in the study and for the conversation to be recorded and used for research purposes. 

\subsubsection{Data analysis}
The interviews were transcribed for analysis. Using the affinity diagramming method, the three researchers independently reviewed and extracted key insights, which were then collaboratively grouped into thematic clusters. Specific quotes were tagged with participant IDs (P1–P11) to preserve source traceability. This process resulted in the identification of four core themes, which are detailed in the next section.

\subsection{Findings from the expert interviews}
We clustered our findings from the interviews with the HR professionals into four themes: (1) challenges in onboarding and social bonding, (2) challenges for communication, (3) collaboration and team roles, and (4) digitization and team development.

\subsubsection{Challenges in onboarding and social bonding}
Participants repeatedly emphasized the emotional gap in onboarding and the erosion of informal social bonds, particularly after the COVID-19 pandemic. Although the cognitive components of introduction to the organization, such as tools and tasks, are largely preserved, companies are increasingly struggling to maintain the relational and emotional aspects that are crucial to building organizational culture.
For example, P1 noted: \textit{“Creating real engagement beyond the job description is difficult”} while P4 reflected on the loss of informal rituals post-COVID: \textit{“We have lost informal team rituals like the Monday morning breakfast meeting.”} The social cost of hybrid work is acknowledged at all organizational levels (P8), and leaders are acutely aware of how important employee participation is to organizational success. However, as P3 observed, it is difficult to foster a sense of \textit{“loyalty and togetherness”} in changing workplace structures. 

In response, several organizations have begun to experiment with new activities. One of the interviewees saw that different departments opted for digital escape rooms as a team building activity (P1). Others turned to creative initiatives, such as co-authoring eBooks or producing short team videos (P11). Traditional methods, including coffee chats, company events, and in-person games, remain widely used to support team bonding (P1, P3, P4, P5, P7, P11). However, as P7 noted, these activities often lack greater participation and do not capture how teams actually function together in practice.

\subsubsection{Challenges for communication}
The interviewees highlighted the evolving communication needs driven by changes in workplace environments. In large organizations, establishing clear rules and structured communication protocols is essential (P5, P6, P9, P10). Participants also differentiated between task-focused communication tools (e.g., quick messaging for execution, P9) and those intended to foster interpersonal connection between team members (P11). Leaders emphasized the need to upskill employees, especially middle management, to navigate digital tools effectively (P4, P7). For all leaders, communication was viewed as a critical skill that requires ongoing development in the face of increasing digitization.

\subsubsection{Collaboration and team roles}
The participants underlined that effective collaboration requires intentional planning, structured tools, and awareness of the roles and responsibilities of the team members (P10, P11, P7). Building collaborative competencies, joint planning, and decision making were mentioned as key areas for employee development (P4, P6). 

Although playful experiences can support social bonding, they were also recognized as effective tools for team training: \textit{ "Games can improve decision making and leadership"} (P4). Ultimately, promoting collaboration involves making team culture explicit, encouraging open discussion of processes, values, and expectations to reduce misunderstandings and performance issues (P6). Participants emphasized the need for clarity of roles and structural awareness in hybrid and cross-functional teams. Ambiguity around roles can hinder collaboration, especially when hierarchical or generational dynamics are involved (P5, P9). Making team structures and expectations explicit was seen as a practical step toward minimizing conflict and improving efficiency. Successful teamwork was also related to a shared understanding of organizational norms and collaborative ownership (P6, P7). 

\subsubsection{Digitization and team development}
Digitization was seen as a strategic challenge and an opportunity in team development. Participants noted the need to redefine traditional practices and integrate digital tools thoughtfully: \textit{“Digitization is no longer optional. It is a strategic concern”} (P7). Rather than simply increasing productivity, digital tools must foster meaningful engagement. As P4 observed, \textit{“We are moving toward agile and lean processes that require empowerment at every level.”} However, passive engagement with digital systems can lead to fatigue (P11), underscoring the need for interactive and emotionally resonant experiences. VR was identified as a promising tool to restore emotional depth to team interactions, but it may also provide a way for employees to try new technologies and become accustomed to them (P5). Immersive meetings are one potential use of VR (P8), but their more powerful potential to build empowerment within organizations has not yet been fully explored (P4). Ultimately, new tools have the potential to deepen conversations about how teams function and how collaboration can be improved (P7).

\subsection{Skills, Requirements, and Features}
\begin{table*}
  \caption{Topics, KSAs, Requirements, and Features based on interview data and literature}
  \label{Requirements}
  \setlength{\extrarowheight}{.5em}
  \begin{tabular}{p{9em}p{12em}p{9em}p{12em}}
    \toprule
    \textbf{Interview Topic} & \textbf{Team Development KSA} & \textbf{Requirement} & \textbf{Design Feature(s)} \\
    \midrule
    T1. Affective elements of team development. (P1, P3, P4, P7) &
    KSA24. Team members participate in shared rituals and affirm team identity. &
    R1. Playful multiplayer experience for relationship development. &
    F1. Narrative-driven joint challenge. \\

    T2. Sense of togetherness and social bonding in teams. (P1, P3, P4, P5, P7, P11) &
    KSA15. Recognize and reinforce the ability to complete tasks. \newline
    KSA19. Belief in the team’s ability to overcome obstacles. &
    R2. Shared, multiplayer goal solvable only through collaboration. &
    F2. Use of metaphors like "crew" to build shared identity. \newline
    F3. Avatars representing team members in virtual space. \\

    T3. Communication rules, structures, and protocols. (P5, P6, P9, P10) &
    KSA7. Communicate openly and supportively. \newline
    KSA8. Use active listening techniques. \newline
    KSA21. Seek input and accept feedback constructively. &
    R3. Users must actively communicate and exchange information. &
    F4. Spatial separation requiring verbal exchange. \\

    T4. Collaboration: joint planning and decision-making. (P4, P6, P7, P10, P11) &
    KSA13. Coordinate and synchronize interdependent tasks. \newline
    KSA17. Offer/request backup to support task success. &
    R4. Scenarios must require coordination and interdependence. &
    F5. Scenarios with information asymmetry requiring exchange. \newline
    F6. Inventory system for collecting shared task items. \\

    T5. Awareness of tasks, roles, and structures. (P5, P6, P7, P9) &
    KSA14. Establish and balance role expectations and workload. \newline
    KSA16. Integrate teammates' contributions into task execution. &
    R5. Scenarios must require role division and clear task distribution. &
    F7. Object exchange for interdependent role actions. \newline
    F8. Spatial separation to reinforce task ownership. \\

    T6. Addressing team culture and processes. (P6, P7, P9) &
    KSA27. Generate alternative strategies and compare impacts. \newline
    KSA28. Adjust strategies based on performance feedback. &
    R6. Include reflection and structured debriefing post-gameplay. &
    F9. Structured reflection handouts for feedback and process review. \\

    \bottomrule
  \end{tabular}
\end{table*}

We systematically mapped the themes emerging from the interviews against established KSAs illustrated in Table \ref{tab:ksa_categories}. We then translated these skills into requirements (specifications) and features (distinctive characteristics) that the system needs to have to support team training. Table \ref{Requirements} provides an overview of the interview topics, KSAs, requirements, and features. 

\subsubsection{Topics and KSAs}
We reviewed the interview data and selected the topics that were most frequently mentioned. Our criteria were that a topic had to be mentioned by at least three participants and focus on team-related skills and abilities. We then reviewed the literature to frame these topics according to established teamwork KSAs. To preserve traceability, we retained the original KSAs numbering shown in Table \ref{tab:ksa_categories}. In total, we identified 12 KSAs addressed in the topics mentioned by the interviewees, covering the following categories: communication (KSA7, KSA8), goal setting (KSA13, KSA14), collective efficacy (KSA15, KSA16, KSA17, KSA19), trust (KSA21), cohesion (KSA24) and reflexivity (KSA27, KSA28). 

While Table \ref{tab:ksa_categories} presents a comprehensive set of 28 teamwork KSAs, our work deliberately focuses on a subset that were selected because they (1) were emphasized by HR experts in our interviews and (2) could be meaningfully enacted within two short, co-located, asymmetric VR scenarios. In particular, KSAs related to communication, coordination, collective efficacy, cohesion, and reflexivity form a foundational layer of teamwork that can be effectively addressed through asymmetric, interdependent interaction.
Several unaddressed KSAs, such as those related to conflict resolution, trust calibration, and performance monitoring, would require dedicated scenarios that rely on different design mechanics, including intentional breakdowns or deliberately adversarial dynamics. We reflect on this choice in the Limitations section.

\subsubsection{Requirements and features}
\highlight{Interactive experiences can foster affective connection and a sense of togetherness by aligning participants around shared goals \cite{tekinbas_rules_2003}. Accordingly, we designed the experience as a multi-player scenario driven by narrative and coordinated challenges (F1), enabling participants to work toward shared outcomes and experience collective accomplishment (KSA24, KSA15, KSA19). We further reinforced a sense of group identity through team-related metaphors such as “crew” (F2). In addition, representing participants through avatars in a shared virtual space can strengthen social bonding, especially when users are co-located and can perceive each other’s actions in real time (F3).}

To train communication and active listening (KSA7, KSA8, KSA21), users need to communicate and exchange information (F4). As we showed in Section 2.2, other playful experiences demonstrated the potential of aymmetry of information, where players must communicate verbally instructions or solutions to each other \cite{harris_asymmetry_2019, harris_beam_2015, goncalves_exploring_2021}.  

Training coordination and task division scenarios (KSA13, KSA14, KSA16, KSA17) can require users to manage interdependency while also dividing roles. For example, physical separation in the digital space means that users do not have access to the same objects or information (F5). Moreover, while an inventory can support the collection of objects, an item exchange function can give players more flexibility in changing task division and roles (F6, F7). Reflection on team processes (KSA27, KSA28) means that users need space during or after sessions to debrief and reflect on their performance \cite{miller_design_2024}. Considering the potential to transfer learning to workplace collaboration, a structured debrief outside the experience can help users reflect on and improve their gameplay strategy (F10). 

\subsection{Experience design}
Based on the above requirements and features, we designed an interactive environment that challenges participants to engage in collaborative problem-solving and develop skills that can be transferred to work settings. We chose to set the scenarios aboard a fictional spaceship stranded in space, to metaphorically create a sense of togetherness, and to symbolize shared objectives (KSA1, KSA2).

The VR application was developed using Unity3D, ARKit, and ARCore, ensuring compatibility with iOS and Android handheld devices such as smartphones and tablets. Although HMDs offer enhanced 3D interaction and greater immersion, we chose handheld devices as the target platform due to their broader accessibility, making them a more practical choice for widespread adoption in corporate training contexts. Multiplayer functionality was implemented through Unity's Netcode. 

The application features custom-designed 3D environments and integrated elements of asymmetric games, puzzle games, and cooperative problem-solving. The users interact with the environment by tapping and pointing at different objects using the touch screen, saving them in their inventory, and exchanging them with the other users. We focused on \textit{ information asymmetry} (through spatial separation) and \textit{ ability asymmetry} (through tools) to reinforce the need to train teamwork KSAs. We also create \textit{desirable difficulties} by not implementing additional awareness cues between players, such as raycasting or screen sharing. However, we used virtual avatars to represent players when located in the same virtual space. 

Finally, since team development is a process in which behaviors and performance improve over time \cite{tuckman_stages_1977, tuckman_developmental_1965}, we propose a progressive design consisting of two scenarios, broadly based on Tuckman's team development stages. Scenario 1 is an introductory task, corresponding to the \textit{forming} stage: it sets the context, the metaphor and the goal of the crew (KSA24, KSA15, KSA19). Scenario 2 is fast paced and increases in complexity and difficulty. This may pose additional challenges and therefore has similarities to the \textit{storming} stage. 

\subsubsection{Scenario 1: \textit{Lost in Space}}
\begin{figure}
  \centering
  \includegraphics[width=\linewidth]{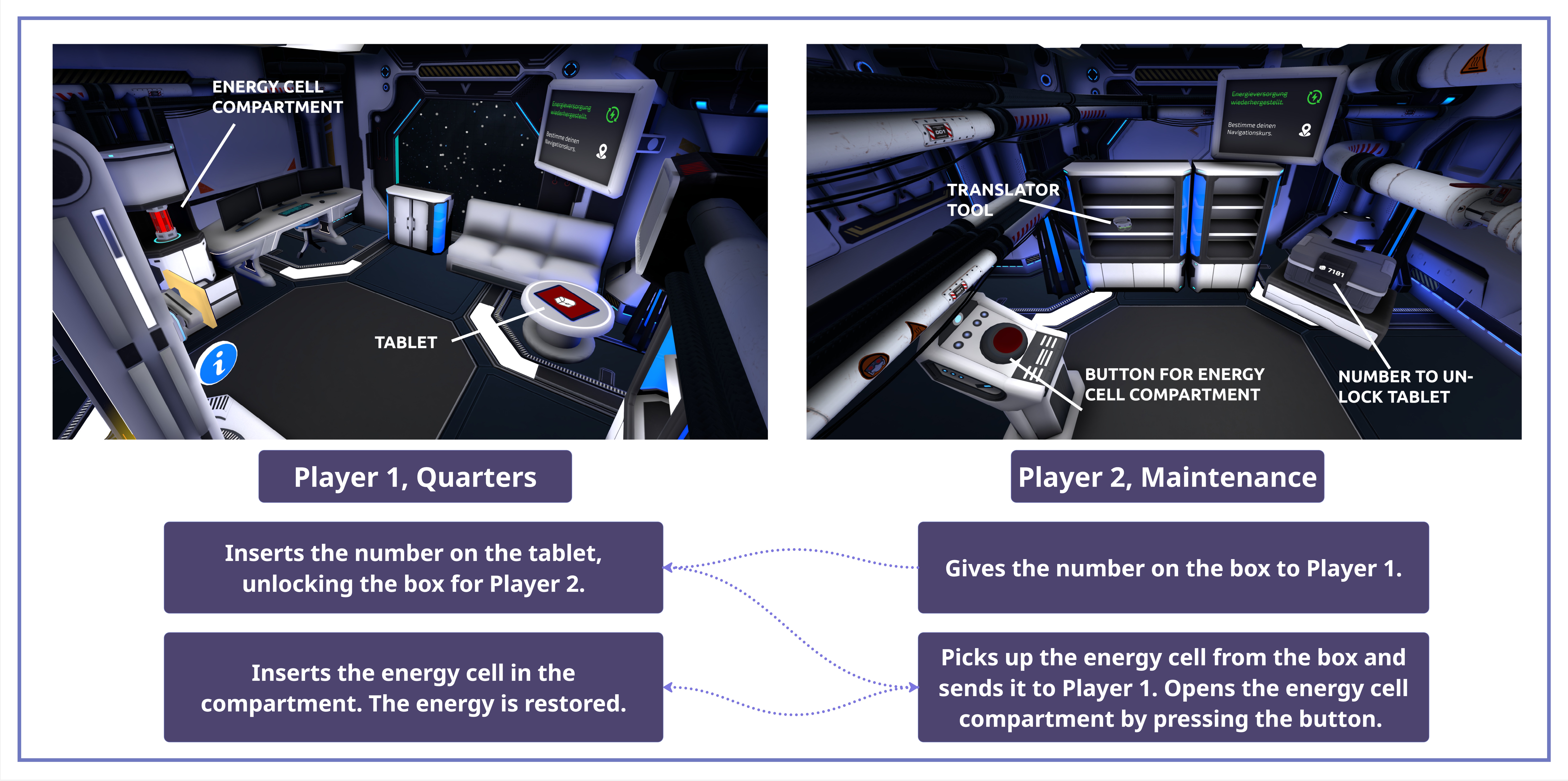}
  \caption{Annotated screenshots of the VR environment, communication and task interdependencies in Task 1 of \textit{Lost in Space}}
  \label{Task1}
  \Description{The figure shows the VR environment of each player, the items available to them and the sequential ordering of tasks.}  
\end{figure}

The first scenario is designed for two players who start in different rooms on the spaceship: Player 1 begins in the Quarters; Player 2 starts in Maintenance (R5, F8). The players cannot change location due to a power malfunction. To successfully solve the scenario, they must perform two tasks: (1) restore the energy supply of the ship and (2) determine the location of the ship in space (R1, F1, F2). In the first task, Player 2 is equipped with a translator and can read instructions on screens, but cannot unlock a box containing an energy cell. Player 1 has access to a tablet that displays a code required by Player 2 but cannot interpret other instructions due to the display of a foreign font (R3-R5, F4-F8).
The task requires sequential collaboration, helping participants clarify goals, share information, and coordinate actions to advance in the game. The experience has no time limit. Figure \ref{Task1} depicts the VR environment and the sequential task distribution in Task 1. Players develop KSAs by solving the tasks:   
\begin{itemize}
   \item Communication and information exchange: Player 2 shares the code with Player 1 to unlock the box. Player 1 describes the energy compartment cell and requests help opening it (KSA7, KSA8, KSA21).
   \item Task interdependence: Player 1 is needed by Player 2 to provide the code. Player 2 retrieves the energy cell from the box and passes it to Player 1. They open the energy cell compartment by pressing the button. Player 1 installs the energy cell in its compartment, allowing the ship's energy systems and the Transporter to become operational (KSA13, KSA17).
   \item Team roles: Each player has a designated role based on their location on the ship (KSA14, KSA16). 
\end{itemize}

\begin{figure}
  \centering
  \includegraphics[width=\linewidth]{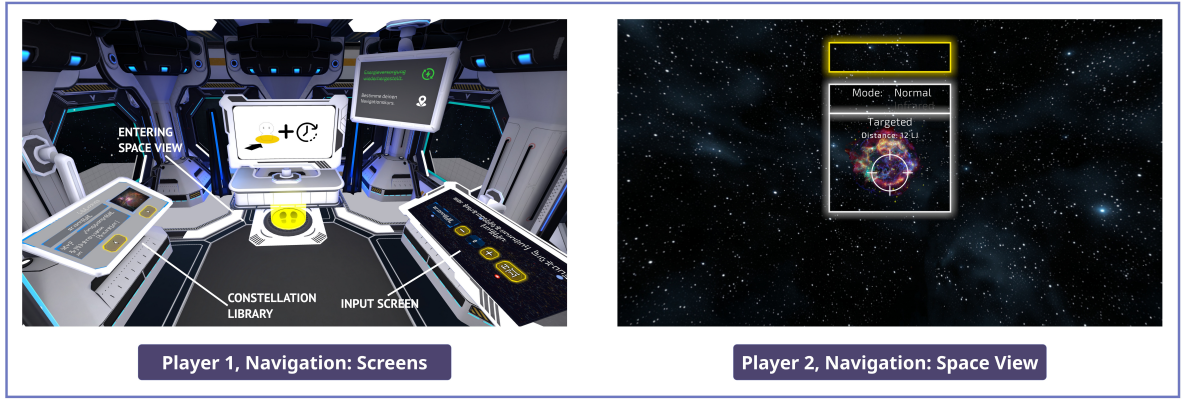}
  \caption{Annotated screenshots of the VR environment, communication and task interdependencies in Task 2 of \textit{Lost in Space}}
  \label{Task2}
  \Description{The figure shows the VR environment of each player and the items available to them.}  
\end{figure}

The second task builds on the foundation of collaboration established in the first. Both players must meet in the Navigation room and work together to triangulate the spaceship’s position using three stars or constellations. They have access to two screens: one for inputting distances and one for displaying constellations. An immersive Space View enables them to observe the surrounding stars and galaxies. Figure \ref{Task2} depicts the VR environment and interactable objects for each player in Task 2. 

Players can also retrieve Infrared binoculars to reveal otherwise invisible constellations. Only one player can enter the Space View at a time, and only one player can view the information on the screens, the player holding the Translator (R4, R5, F5, F6, F8). However, players have more flexibility in coordinating their actions and switching roles: they can negotiate who is holding the Translator and who is holding the Infrared and working with the Space View. This task also targets teamwork KSAs: 
\begin{itemize}
   \item Communication and information exchange: The player using the screens must describe to the player in the Space View what the constellations look like (KSA7, KSA8, KSA21). 
   \item Task interdependence: The player in the Space View can only see the stars and their distances, but does not know their names. The Player looking at the screens knows which star constellations need to be inputted and must describe them to the Player in the Space View (KSA13, KSA17, KSA14, KSA16). 
   \item Team roles: Players negotiate roles: who is holding Infrared and Translator (KSA14). Players can exchange tools and location (KSA16, KSA28). 
\end{itemize}

\subsubsection{Scenario 2: "The Journey"}
\begin{figure}
  \centering
  \includegraphics[width=\linewidth]{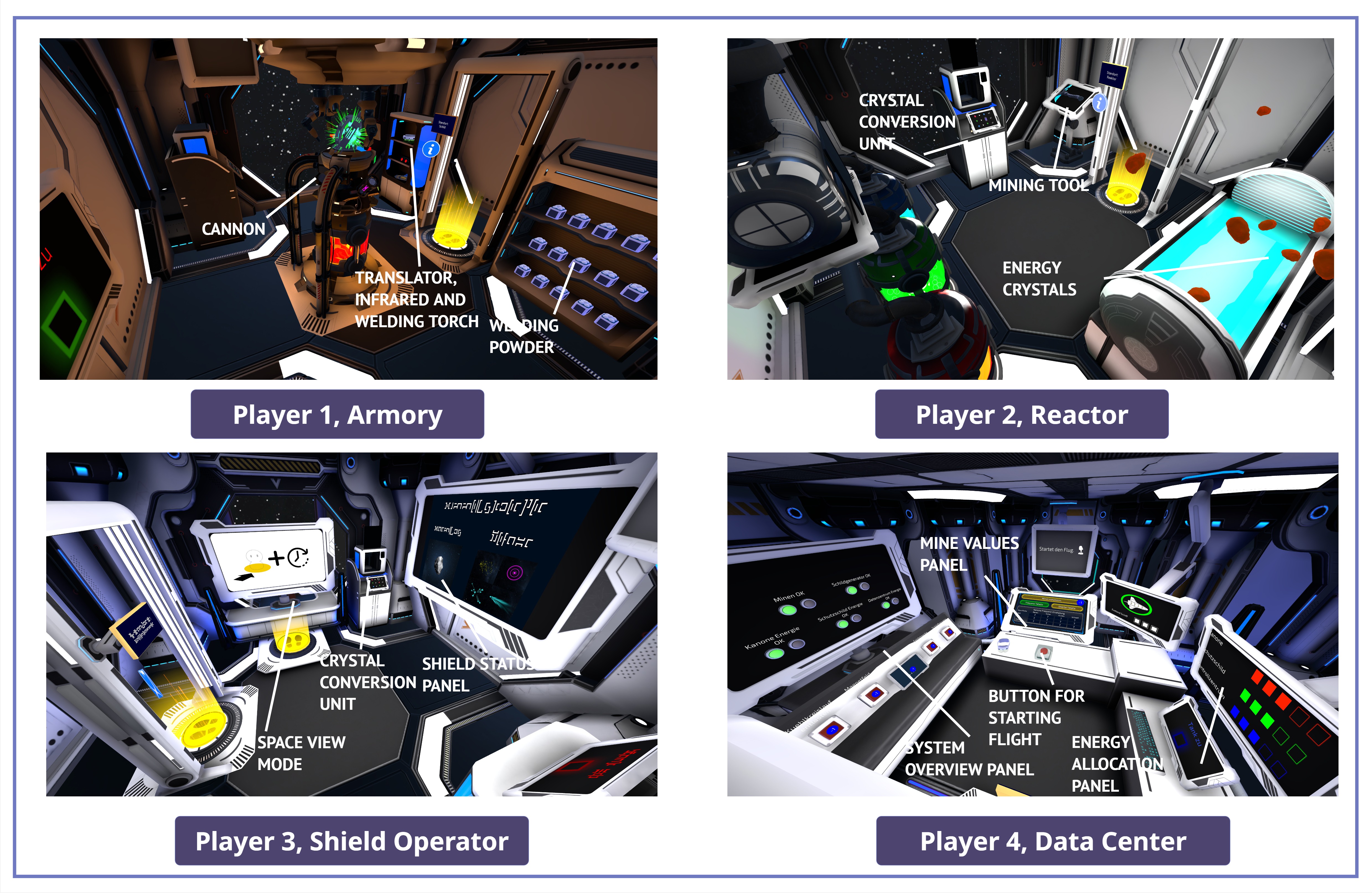}
  \caption{Annotated screenshots of the VR environment and objects in Scenario 2 - \textit{The Journey}}
  \label{Task3}
  \Description{The figure shows the VR environment of each player and the items available to them.}  
\end{figure}

In \textit{The Journey}, a four-player team is tasked with navigating a minefield while managing resources and responding to errors, which requires tight coordination. Each player is assigned to a different room from the beginning. Different tasks must be performed simultaneously, and each team member collaborates with the other players for specific needs. For example, Player 1 is assigned to the Armory room and must configure and operate the laser cannon to destroy mines. They must adjust the color, phase, and pulse pattern of the cannon waves to match the characteristics of each mine, as determined by Player 4. Player 4, assigned to the Data Center, is in charge of managing the mines database, energy allocation, and error correction. They coordinate closely with all players, providing oversight and instructions. Figure \ref{Task3} visually illustrates the VR environment of each player and the items available to them. 

\textit{The Journey} is a timed challenge: players must navigate the minefield for 20 minutes without losing shields. The timer is not visible to players. The scenario has a higher level of difficulty: most teams would not be able to successfully complete the mission on the first attempt. This enables more elaborate reflections on adapting tasks and processes, as part of the debrief. The change in team size from two to four players also impacted the complexity of coordination. 

Table \ref{Scenario2} further details the location, role, and task of each player. By assigning different roles and emphasizing interdependence, this experience simulates scenarios where individuals must perform under stress while tightly coordinating. In doing so, they practice teamwork KSAs:
\begin{itemize}
   \item Communication and information exchange: Players need specific information from each other: mine frequency patterns, shield status updates, error management, energy needs. Players must provide input when needed, provide support, and quickly integrate feedback (KSA8, KSA17, KSA16). 
   \item Task interdependence: The tasks require players to coordinate closely with each other: providing support, relaying information, passing objects (i.e. energy crystals), and clarifying roles and inputs (KSA13, KSA17, KSA14, KSA16). Players often require the support of their peers, as they must manage errors in parallel to other tasks. 
   \item Team roles: Players are assigned roles through spatial allocation to the different rooms. However, they are able to move to a different room and change roles (KSA14). 
\end{itemize}

\begin{table}
  \caption{Roles and interdependencies in Scenario 2}
  \label{Scenario2}
  \setlength{\extrarowheight}{.5em}
  \begin{tabular}{p{4em}p{4em}p{7em}p{12em}p{12em}}
    \toprule
    Player & Room & Role & Task & Collaboration points\\
    \midrule
    Player 1 & Armory & Operates the cannon. & Adjust cannon wave, color and phase pattern to mines. & Relies on Player 4 for information about mines. \\
    Player 2 & Reactor & Converts crystals into energy. & Identifies and processes crystals into energy. & Player 4: crystal allocation. Player 3: shield energy. Player 1: cannon requirements.  \\
    Player 3 & Shield & Manages the shield. & Communicates shield status and manages errors caused by shield failure. & Alerts Players 2 and 4 to reallocate energy. \\
    Player 4 & Data center & Manages the database and energy. & Relay mine values to Player 1, allocate energy, handle errors. & Coordinates closely with all players. \\
    \bottomrule
  \end{tabular}
\end{table}

\subsubsection{Facilitation and workshop format}
To support reflexivity (KSA27, KSA28) we anchored the experiences in a workshop format, where each play session is followed by a reflection round. We created handouts for pairs and groups to use for the debrief after playing together. For scenario 1, \textit{Lost in Space}, the reflection prompts were:
\begin{itemize}
    \item What were our key moments in solving the game? 
    \item How did we coordinate with each other and agree on goals and procedures? 
    \item How did we communicate with each other? 
    \item How do we want to adapt our approach and communication in the next round? 
\end{itemize}

For scenario 2, \textit{The Journey}, the prompts were as follows: 
\begin{itemize}
    \item How did we start as a group in the new constellation?  
    \item How did communication change when switching from pairs to groups? 
    \item How would we change our approach in the next round? 
\end{itemize}

\section{Study design}
To address \textbf{RQ2}, we conducted an exploratory study to investigate \textbf{how the design of the playful experience supports the development of teamwork KSAs}. 

\subsection{Setup}
We conducted a three-hour workshop with 16 participants, structured as follows: 
\begin{enumerate}
    \item Introduction (15 minutes). 
    \item Interactive tutorial played individually (10 minutes). 
    \item Play session "Lost in Space" (15 minutes), questionnaires (10 minutes), team reflection (15 minutes) and group discussion (20 minutes). 
    \item Play session "The Journey" (30 minutes), questionnaires (10 minutes), team reflection (15 minutes) and group discussion (20 minutes).
    \item Final feedback (10 minutes). 
\end{enumerate}

The interactive tutorial aimed to familiarize participants with basic user interactions. Three researchers ran the session: two for onboarding and discussion and one for handling the devices. Participants were provided with iOS devices: 14 iPads (7 iPad mini, 7 iPad Pro) and two iPhones. We used an independent network through a 5G router. 

\subsection{Participants}
The 16 participants were university students (M=13, F=3), enrolled in a multimedia technology program at a local university. We opted for recruiting students, as they represent future employees and apprentices for whom team training is increasingly relevant \cite{noauthor_future_2025}. Moreover, for younger participants, immersive and interactive formats such as VR may offer higher engagement than traditional presentation-based training.

The youngest participant was 22 years old, the oldest was 29, the median age was 24. Although familiarity with playful experiences was not required for the study, we asked participants about their familiarity with games and escape rooms with the two questions: "How much experience do you have with playing games / escape room games?". We measured both on a seven-point Likert scale (1 ... "no experience at all", 7 ... "a lot of experience"). As escape games are popular problem-solving games \cite{shakeri_escaping_2017, warmelink_amelio_2017}, we considered it to be a good indicator that resembles elements of our scenarios. The mean value for the game experience was 5.56 and the escape room experience was 4.25, making the group relatively experienced with games. 

\begin{table*}
  \caption{Questionnaire items}
    \label{Questionnaire items}
    \setlength{\extrarowheight}{.5em}
  \begin{tabular}
  {p{20em}p{22em}}
    \toprule
    Questionnaire item & Answer Options\\
    \midrule
       How closely did you work together with your partner to solve the room? & We worked independently all the time (1) - We worked together all the time (7) \\ 
       
       How closely did you observe the work your partner during the gameplay? & I had no idea what my partner was doing (1) - I was fully aware of what my partner was doing (7) \\ 
       
       Did you explicitly divide up the tasks between you? & Yes / No \\ 
       
       Did you and your partner work effectively as a team to solve the room? & Very ineffectively (1) - Very effectively (7)\\
       
       How much did you personally contribute to solving the tasks? & I contributed the least (1) - I contributed the most (7)\\
       
       How satisfied are you with your work in the pair to solve the room? & Very dissatisfied (1) - Very satisfied (7) \\
         \bottomrule
\end{tabular}
\end{table*}

\subsection{Instruments}
To measure collaboration, we used the questionnaire in Table~\ref{Questionnaire items}, consisting of items adapted from Jakobsen \cite{jakobsen_up_2014}. All elements were measured on a seven-point Likert scale except task division, which was assessed using a dichotomous self-report item (“Did you explicitly divide the tasks between you?”; Yes/No, see Table 5). We video-recorded the session using 3 Go Pro cameras, capturing the group formations and interactions from different angles. We audio recorded the group discussions after each reflection round, where teams shared their impressions based on the handout questions. 

\subsection{Data analysis}
\subsubsection{Questionnaires}
For the questionnaire analysis, we used SPSS Statistics version 29. We calculated the frequencies for the different answers, including for the dichotomous question. 

\subsubsection{Audio recordings}
The audio recordings of the two group discussions were transcribed using automated tools. We used qualitative content analysis \cite{drisko_content_2016}, which initially resulted in 88 open-coded segments. From the initial open codes, we defined eight descriptive categories that represented the input of the participants: (1) awareness, (2) clues, (3) puzzles, (4) communication, (5) exploration and orientation, (6) achievement of goals, (7) division of tasks, and (8) tools. After a summative check against our research question, we clustered these into three categories, which are described in the Findings section. 
\begin{enumerate}
    \item \textbf{Communication and information exchange}: Communication behaviors, strategy sharing, cue interpretation, and coordination through dialogue. Includes the subcategories: communication, awareness, and clues. 
    \item \textbf{Spatial separation and navigation}: How physical separation in virtual space shapes interaction and navigation. Includes the subcategories: exploration and orientation, puzzles.
    \item \textbf{Interdependency and role distribution}: Division of labor, interdependent actions, and alignment of goals. Includes the categories task division, goal achievement, tools. 
\end{enumerate}

\subsubsection{Video recordings}

\begin{figure}[h]
  \centering
  \includegraphics[width=\linewidth]{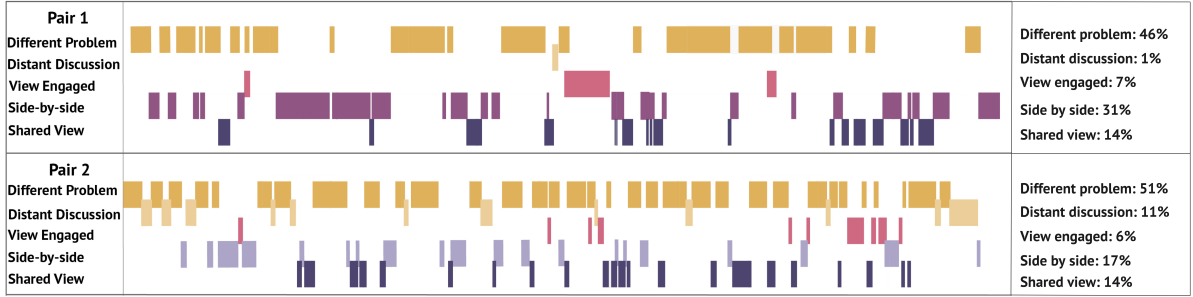}  
  \caption{Codeline examples: Pair 1 and Pair 2 including percentage calculations for each coupling style}
  \label{Fig:Codelineexample}
  \Description{The figure shows the codelines for two of the pairs including percentage calculations.}
\end{figure}

We used MaxQDA 24 for video coding and the \textit{Domino} coding framework \cite{neumayr_domino_2018} as a starting point (Table~\ref{tab:video_framework1}). We watched the videos several times, each time focusing on a particular pair or group. For each pair, we first deductively coded the data using time-based segments, capturing changes in coordination style as participants interacted with the system and with each other. Each coded segment began when participants entered a specific coordination style and ended when this style changed, resulting in continuous codelines for each pair and group that reflect switches between different coordination modes (see Figure \ref{Fig:Codelineexample}).

After coding the first four pairs, two researchers discussed the codelines and observations, adding four codes not captured in previous studies: SSD, UD, DD, and DO. These codes indicate behaviors that are not reflected in codes from previous work (e.g. \cite{neumayr_domino_2018, wells_collabar_2020}). For example, Active Discussion refers to participants who interact with each other, without interacting with the system. Due to the asymmetric nature of the tasks, in our study, participants frequently worked individually while discussing side by side (SSD) or from farther away (DD). We then recoded the data from the first four pairs with additional new codes and processed the remaining four pairs. We used the same procedure to code the video data from the second round of gameplay, additionally coding the subgroup formations: pairs, triad, or quartet. 

To quantify collaboration patterns, we aggregated the durations of all time-based segments assigned to a given code and calculated the percentage of total gameplay time spent in each coordination style for each pair or group.

\begin{table}
\caption{Collaboration modes: coding framework for video data adapted from previous work. Bold font stands for new codes we introduced during the analysis}
\label{tab:video_framework1}
\begin{tabular}{p{10em} p{32em}}
\toprule
\textbf{Code} & \textbf{Description} \\
\midrule

\multicolumn{2}{l}{\textbf{Tightly Coupled}} \\
\midrule
DISC: Active Discussion & Two or more participants actively discuss, with minimal system interaction. \\
SSP: Same Problem & Participants divide effort to solve the same problem, often dealing with different information and tools. \\
VE: View Engaged & One participant is actively working while the other watches without interacting. \\
\textbf{SSD}: Side-by-Side & Participants work individually while discussing side by side. \\
SV: Shared View & Participants look at the same physical display, usually one player’s tablet. \\
\textbf{UD}: Use Device & One participant interacts directly with another's tablet. \\

\midrule
\multicolumn{2}{l}{\textbf{Loosely Coupled}} \\
\midrule
DP: Different Problem & Participants focus on different sub-tasks or problems. \\
\textbf{DD}: Distant Discussion & Participants converse or request information while physically distant or moving. \\
\textbf{DO}: Distant Observation & One participant observes from afar without engaging or interfering. \\
D: Disengaged & A participant is passively watching or not engaged in the task. \\
\bottomrule
\end{tabular}
\end{table}

\section{Findings}
We present the findings of the questionnaire, the reflections resulting from the group discussions, and the findings of the video data on coupling styles. 

\begin{figure}[h]
  \centering
  \includegraphics[width=\linewidth]{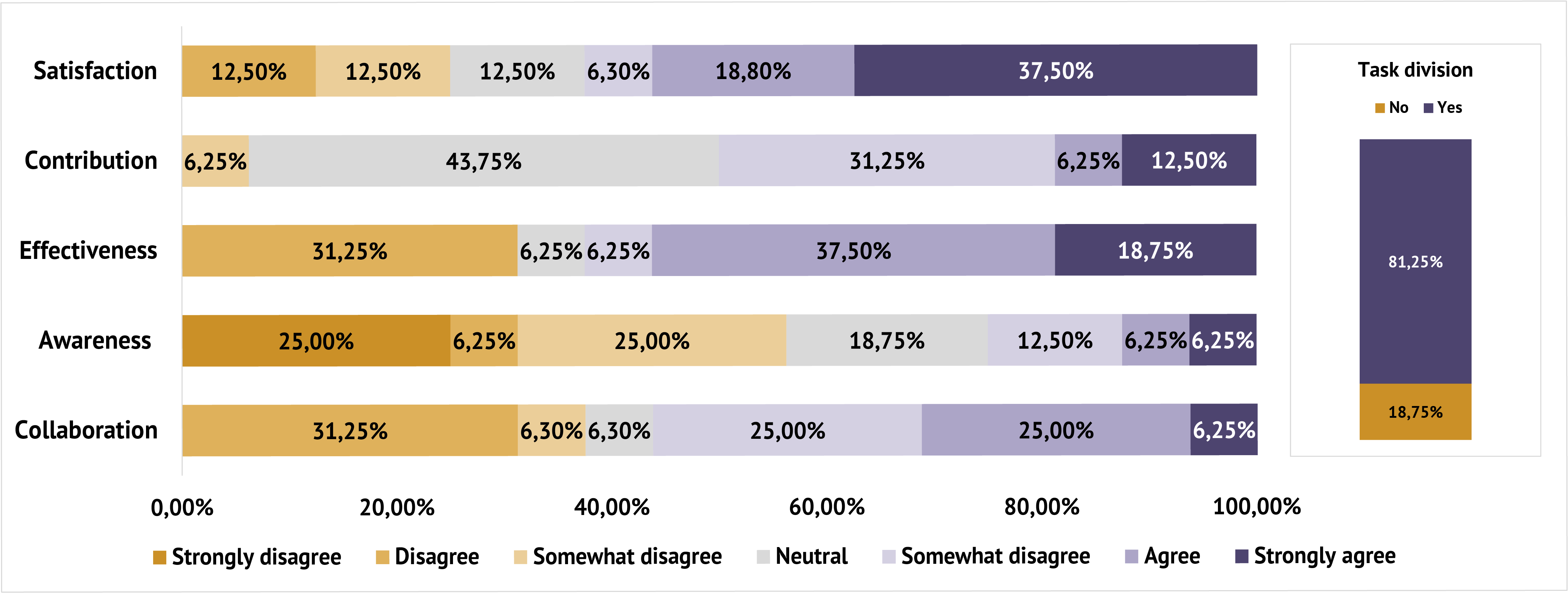}
  \caption{Questionnaire results: "Lost in Space"}
  \label{Fig:Questionnaire1}
  \Description{The figure shows the responses of participants in percentages, for the 7-point Likert scale.}
\end{figure}

\subsection{Questionnaire results}
\begin{figure}
  \centering
  \includegraphics[width=\linewidth]{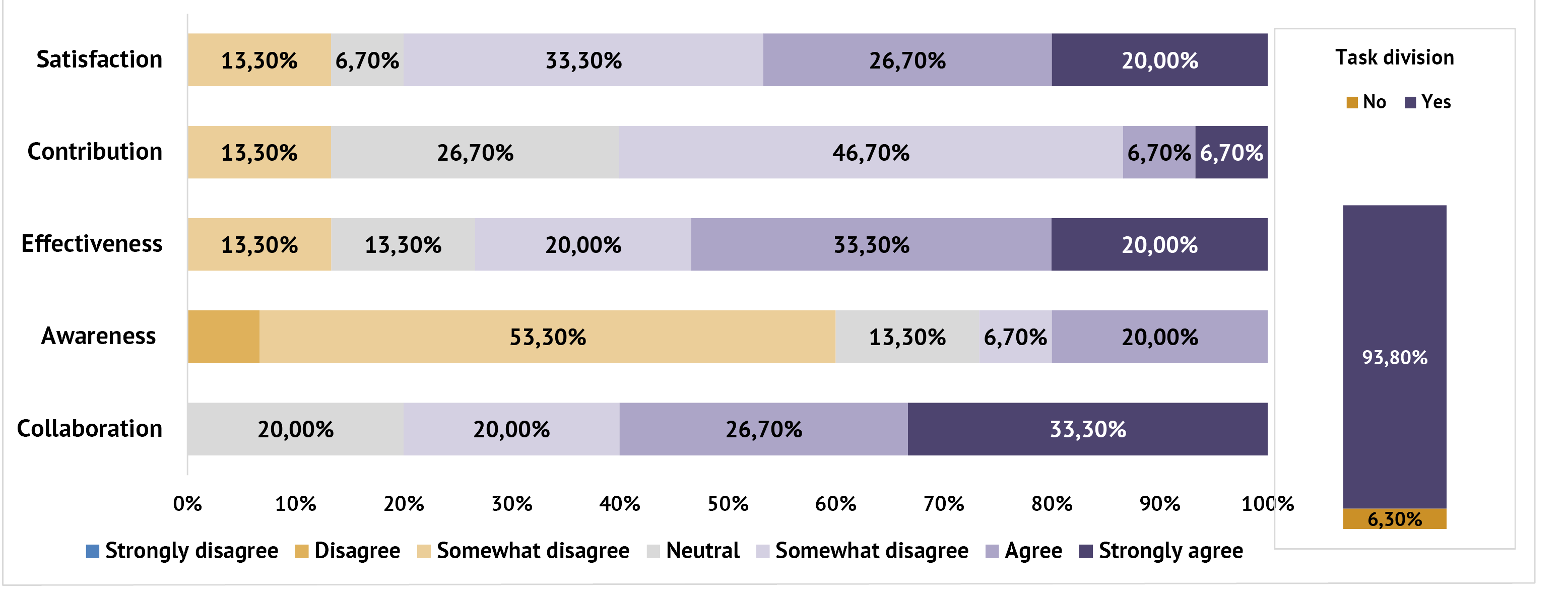}
  \caption{Questionnaire results: "The Journey"}
  \label{Fig:Questionnaire2}
  \Description{The figure shows the responses of participants in percentages, for the 7-point Likert scale.}
\end{figure}

Figures \ref{Fig:Questionnaire1} and \ref{Fig:Questionnaire2} show the participants' responses to the questionnaire, after each round of gameplay. As shown in Table~\ref{Questionnaire items}, the items measured the perception of collaboration of the participants, measured on a seven-point Likert scale, with the exception of the task division item, which was measured dichotomously (see the questionnaire items in Table 5). In \textit{Lost in Space}, participants felt largely effective and structured, but some struggled with awareness and communication. The items most positively ranked are \textit{Satisfaction}, \textit{Effectiveness}, and \textit{Collaboration} but the negative responses suggest that perceptions were split. The variance in answers may reflect early-stage coordination challenges or the need for clearer roles. The item with the highest negative rating was \textit{Awareness}, where the players indicated that they were not always aware of the actions of their partners. 

In the second round of gameplay, the responses improved for most of the items. 60\% of the participants agreed or strongly agreed that they worked closely with their teammates. \textit{Satisfaction}, \textit{Contribution} and \textit{Effectiveness} were all positively ranked while \textit{Awareness} remained the item with the most negative responses. 

Comparing the results between the first and second experience, we notice an overall improvement in perceived collaboration. The answers to the second round are more cohesive, demonstrate improved collaboration, and high levels of task division. This suggests that group reconfiguration, progressive game design, and reflection helped improve teamwork skills. \textit{Awareness} remained the most mixed category, due to our choice of asymmetric design: users were intentionally separated and had to communicate verbally, without the support of other tools. The findings suggest that the participants became more coordinated and confident in their collaboration after initial exposure and reflection.

\subsection{Findings from the group discussion}

\subsubsection{Communication and information exchange}
In both rounds of gameplay, the participants emphasized the importance and limitations of verbal communication to build shared understanding. Most of the communication was brief, focused on the task, and occurred in bursts: \textit{"I needed this", "Okay, I have this"} (P4). This style, described by one participant as "gamer communication", was efficient and factual. The players also used more detailed descriptions, especially when the next steps were unclear or they needed assistance. 

Information exchange was critical to solving the puzzles, particularly when the clues were distributed asymmetrically between players. In \textit{Lost in Space}, successful pairs established early protocols for sharing discoveries. Others had not reflected on it before the debrief: \textit{"We did not have a clear system"} (P1). One pair relied on each other not just for exchanging information, but also to recall it: \textit{"I said to him: Remember those numbers because I will just try all of them, one after another, when I go back to the interface with the numbers"} (P6). However, others struggled to maintain mutual awareness, leading to duplicated actions or confusion.

Communication was deemed more difficult in \textit{The Journey}, where multiple people talking at once interfered with each other's information: \textit{If you really needed something from the others, you needed to make yourself heard”} (P1). This was especially difficult when they had to communicate with a person or subgroup while also coordinating with the rest of the team: \textit{"I was trying to explain something, but at the same time someone else was talking"} (P7). As a result, players were more attentive to their team members. Pair 7 exchanged the least amount of information: \textit{"I didn't even know that he had some tools (...), I didn't know that the infrared tool existed”} (P3). 

\subsubsection{Spatial separation and navigation}
The spatial design of the experience significantly influenced the collaboration. In the early stages, participants engaged in open-ended exploration to orient themselves: \textit{"We first explored our rooms - just touched everything to see what worked"} (P5).

This spatial separation often led to temporary confusion about the structure of the environment and the position of each player within it: \textit{I didn’t know which side I was supposed to be on”} (P2). Especially in the first seconds of \textit{Lost in Space}, it took players some time to realize that they were not in the same virtual space, while describing their environment, their actions, and their consequences. For most players, the realization came when one player turned on the light in their room and the other player reported no change on their end.

In \textit{The Journey}, the challenge of mutual awareness was amplified by the complexity of the layout and the connectivity between the rooms. However, this spatial separation achieved its goal: making verbal communication between players a necessity. One participant noted how communication varied in \textit{The Journey}, based on the proximity between rooms and tasks: \textit{“Between the weapons and the information room, communication was tight. With others, it felt more disconnected”} (P4). For most of the players, it was exciting to switch to the second task in \textit{Lost in Space}. It was one of the few moments that players were in the same digital space at the same time. For the rest of the experience, players usually stuck with the rooms and tools they originally picked up. Only one participant navigated beyond his original position in \textit{The Journey}: \textit{"I switched once to look at something”} (P8). In this case, spatial mobility was used as a workaround for communication or coordination gaps.

\subsubsection{Interdependency and role distribution}
The participants used various strategies for task division. In \textit{Lost in Space}, roles were often determined by initial interaction with the tool. For example, the first person to use the Translator typically retained it for the entire session: \textit{“We always thought that’s how it works. We never really thought about it: that one person could have both”} (P7). This approach led to fixed roles, but also created imbalances: in one team, the player without access to tools was disengaged for 26\% of the play session (Pair 7). In the discussion, the player elaborated that they proceeded in this task as if playing alone: "You can do it on your own: I read, I find, I use the infrared, and then go back" (P4).

The second task focused on triangulating the spaceship's position, which gave participants more flexibility in how they shared roles. For example, participants switched using the Space View when one of the players could not find a constellation: \textit{"He just did not find one sun or planet, and after he gave me the tool, I found it"} (P3). For some participants, the perceived scope of collaboration in \textit{Lost in Space} was limited. In the post-session discussion, two players, who noted that they had prior experience in complex collaborative games, described the tasks as relatively straightforward, not requiring excessive amounts of coordination: “I think it was really straightforward” (P6).

In \textit{The Journey}, the interdependency increased in complexity. Participants worked more closely with some teammates than others, often due to overlap in tasks. The pace of exchange and collaboration was much more intense: \textit{"I tried to do my task quickly"} (P3), \textit{"the shield person would always need almost all the resources and always said: 'give me more'"} (P4). In the debrief, participants reflected on their task division and how they could improve their work in a future round: \textit{"One person operating the shields could also help do some of the resource gathering because operating the shields was basically the task with the least amount of effort"} (P6). They also reflected on how the task allocation between the four players resembles roles that are common in teams: \textit{"Command Center is probably Team Lead. Resource generation ... someone who is independent. So, an external person doing their own thing and bringing them to the team"} P7. Moreover, they realized that occasionally they missed requests for assistance from other players \textit{"because everyone was under stress doing their task"} (P8). This served as a reminder that \textit{"if you are stressed, you sometimes lose people"} (P9). 

Despite the added challenge, the participants saw the second session as an opportunity to refine their coordination. No group successfully completed the minefield challenge. In the case of one group, this was due to a malfunction of one of the devices. This created mixed feelings, as some players who were disappointed to not have successfully completed the experience. In the group discussion, participants reflected on the importance of emotions during teamwork and acceptance: \textit{"how do I deal with the fact that I cannot finish things, or I lose, or it's not in the quality I wanted to have it"} (P9). Despite this, all teams expressed optimism about improving performance in a future round.

\subsection{Findings from the video data}

\subsubsection{Coupling styles in \textit{Lost in Space}}

\begin{figure}
  \centering
  \includegraphics[width=\linewidth]{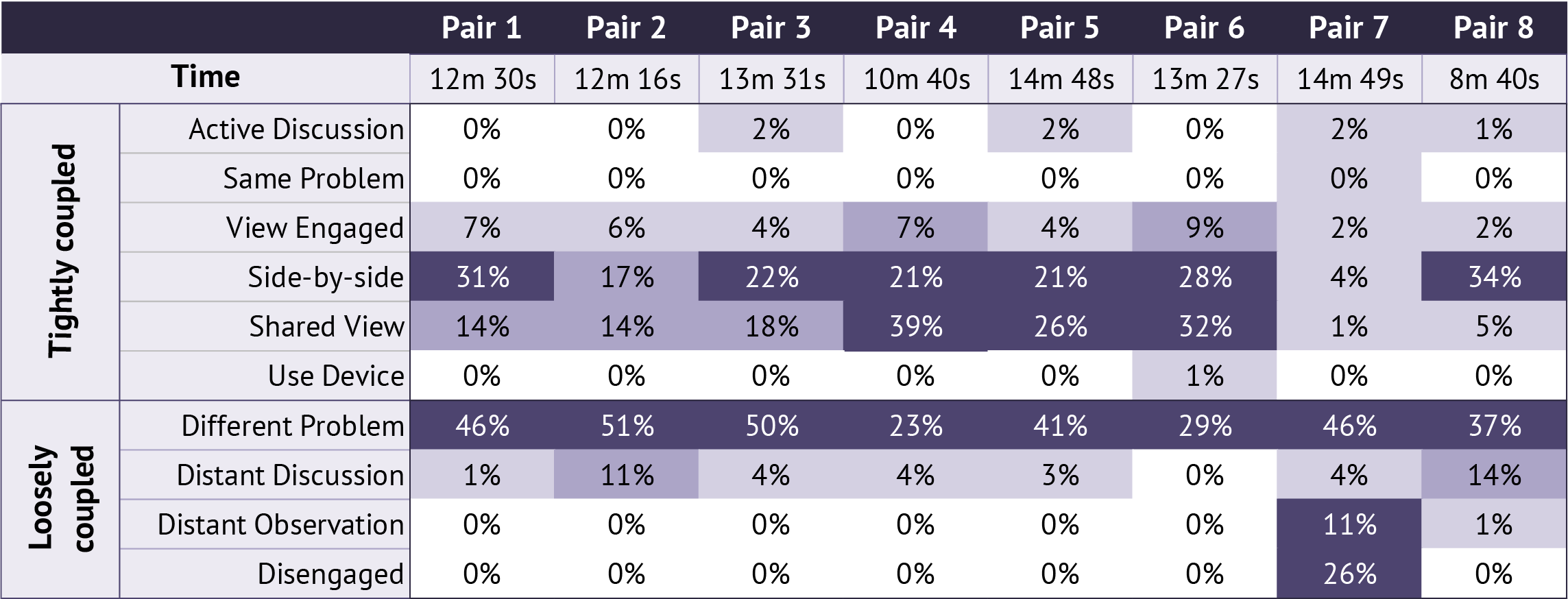}
  \caption{Code Percentages \textit{Lost in Space}. Percentages indicate the proportion of total gameplay time spent in each coordination style per pair}
  \label{CodePercentagesPairs}
  \Description{The figure shows the timeline of the coupling styles for each pair.}  
\end{figure}

Figure \ref{CodePercentagesPairs} shows the distribution of the coordination styles for each pair in \textit{Lost in Space}, expressed as the percentage of the total time spent playing. These percentages were calculated by aggregating the durations of all time-based video segments coded with a given coordination style from Table \ref{tab:video_framework1}. We then divided this sum by the total coded gameplay time for the respective pair. As can be seen in the figure, pairs 1, 2, 3, and 5 spent about half of their time in loose coordination, especially in \textit{Different Problem} mode, and the other half in tight coordination. These patterns suggest a fluid style, where participants alternated between independent work and active coordination, particularly through screen sharing and side-by-side formations. From this selection of pairs, Group 2 spent the least amount of time in tight coordination. 

Pair 4 was the team that spent the most time \textit{Sharing the View}. The most similar behavior is that of Pair 6, who also spent about a third of their time in loose coordination with the remaining distributed between different types of tight coordination. This coordination style frequently coincided with real-time verbal exchange: when participants showed their screens to each other, they typically continued to talk and negotiate the next steps. However, both Pairs 4 and 6  increased the amount of time spent sharing the screen view during Task 2. \highlight{This may indicate that the visual complexity made it easier to show information rather than explain it.}

Pairs 7 and 8 distinguish themselves through different strategies. Pair 7, as noted in the previous sections, struggled the most. After solving the first task, one of the players held both tools (Translator and Infrared) and proceeded to solve the navigation task alone. This rendered the second player unable to contribute, leading them to disengage from the second task. 

Pair 8 completed the experience the fastest, in only 8 minutes and 40 seconds.The completion time is an indication that their collaboration was highly effective. Compared to other teams, they also spent the most time in verbal exchange, even when they were farther away from each other. In contrast, they only shared their views on their tablets in 4,5\% of the time. This suggests that Pair 8 was comfortable exchanging information verbally, did not need to compensate through visual explanations, and communicated efficiently. 

\subsubsection{Coupling styles in \textit{The Journey}}

\begin{figure}
  \centering
  \includegraphics[width=\linewidth]{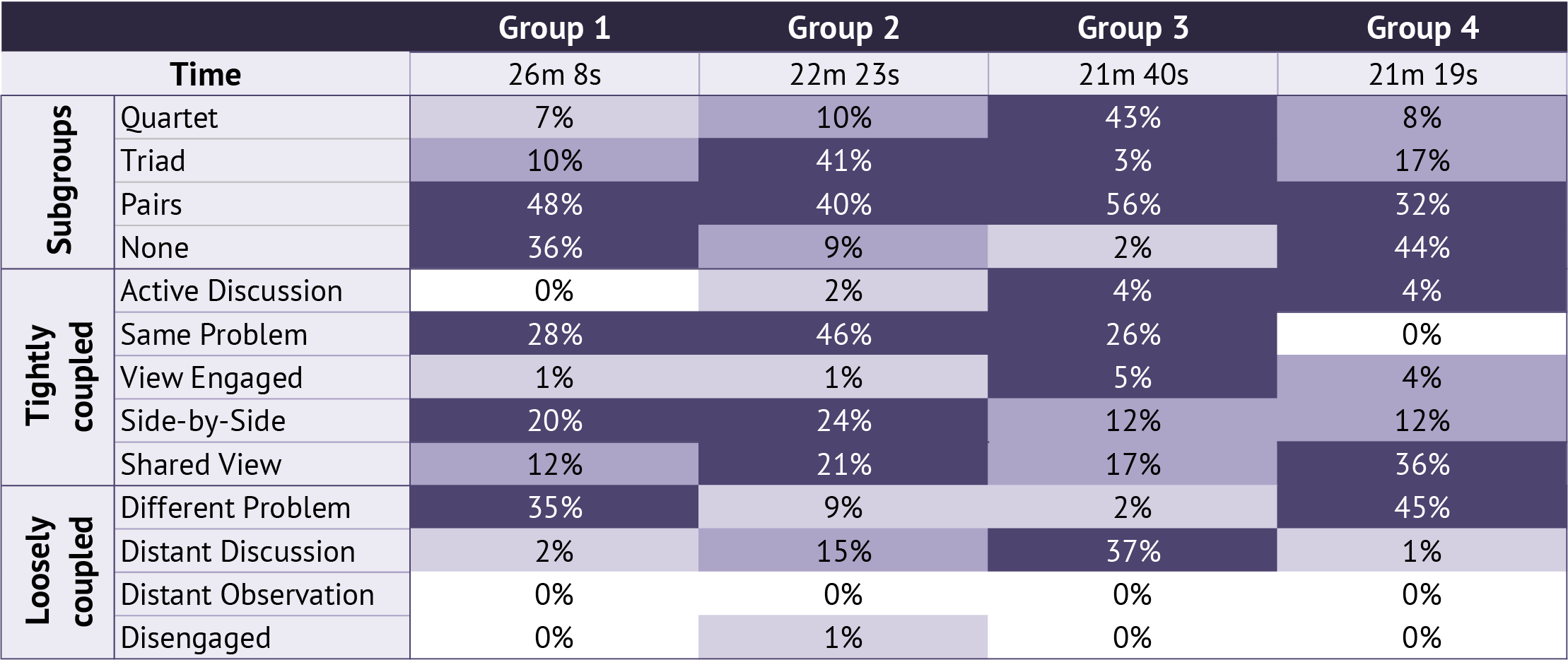}
  \caption{Code Percentages \textit{The Journey}. Percentages indicate the proportion of total gameplay time during which a given coordination style was observed; due to overlapping subgroup activity, totals may exceed 100\%.}
  \label{CodePercentagesGroups}
  \Description{The figure shows the timeline of the coupling styles for each pair.}  
\end{figure}

Figure \ref{CodePercentagesGroups} presents the distribution of the coordination styles for each group in \textit{The Journey} as percentages of the total game time. The percentages were computed by adding the durations of all time-based segments assigned to each coordination style and normalizing them by the total duration of the game of the respective group. Because different team members or subgroups could simultaneously engage in different coordination styles, combined percentages for tightly coupled and loosely coupled modes may exceed 100\%.

Group 1 maintained a fairly balanced coordination profile. They spent 35\% of their time in \textit{Different Problem} mode, indicating individual task focus, but also engaged in tight coordination. This suggests a dynamic collaboration style that alternates between independent contributions and shared focus. Since \textit{The Journey} was a timed challenge, a longer duration spent in the experience means that the teams were actually closer to success. As can be seen in the table, Group 1 spent the longest amount of time playing and therefore was the closest to succeeding.  

Group 2 exhibited the greatest emphasis on tight coupling, with more than 90\% of their time spent in tightly coordinated modes. This group maintained relatively stable subgroup structures, with 41\% of their time in triads and 40\% in pairs, suggesting a collaborative and structured dynamic. However, they also had the highest percentage of time in \textit{Distant Discussion}. Note that in the case of groups, the total coded time in tightly and loosely coupled formations may exceed 100\%, especially if different team members were using different coordination strategies simultaneously. 

Group 3 showed high levels of coordination as a group of four. Their coupling profile included both tightly coupled modes—\textit{Shared View} and \textit{Same Problem} —and a substantial amount of \textit{Distant Discussion}. This suggests that their interaction style relied more on verbal exchanges than continuous joint focus.

Finally, Group 4 relied primarily on loosely coupled collaboration. They spent 45\% of their time in \textit{Different Problem} mode and 44\% without a clearly defined subgroup, indicating a preference for independent work. The tight coordination was more limited compared to the other groups. They seem to have functioned more as parallel contributors than as a fully integrated team. The team had the shortest duration of gameplay, which meant they were farthest from successfully completing the experience. 

\section{Discussion}
We answer our research questions and provide design recommendations for designers, researchers, and practitioners who develop immersive experiences that aim to train teamwork KSAs. 

\subsection{RQ1: Game design elements that support team training}
We evaluate how well our design assumptions from Section ~\ref{sec:designing_vr} held up in practice on the basis of findings from our study. We discuss each design element described in Table \ref{Requirements}. 

\subsubsection{F1: Narrative-Driven Challenge}
Following game design principles \cite{tekinbas_rules_2003}, we used a narrative-driven challenge to increase engagement, shared goals, and emotional involvement. Participants reported being invested in the mission and often used narrative language when describing the experience. However, they were also relatively emotionally restrained, with one exception: in \textit{The Journey}, some members of Group 1 were disappointed to have not completed the mission due to the malfunction of one of the devices. We noticed that the participants were generally engaged in the gameplay, with the notable exception of Pair 7. In their case, this was due to bottlenecks in coordination and communication.

\subsubsection{F2: Use of crew metaphors}
Setting up the experience onboard a spaceship allowed us to foster team identity and collective roles \cite{sottilare_designing_2018}. We found that this metaphor resonated with most of the participants, particularly in \textit{The Journey} , and the players seemed to have internalized their roles quickly. Participants were also familiar with science fiction and, therefore, were quick to assimilate the metaphors. However, this metaphor needs to be further tested with more heterogeneous user groups that may not be equally familiar with this language. 

\subsubsection{F3: Avatars}
In the group discussions, participants did not specifically mention avatars as a means to achieve mutual awareness. Therefore, their impact seems to have been limited, as most participants had already realized that they were not in the same virtual space long before they met in the Navigation room. Given our scenario design, verbal cues were more dominant in coordination. Therefore, this feature requires further testing in scenarios where the virtual co-location is especially relevant.

\subsubsection{F4: Verbal exchange}
The need for verbal exchange was an integral part of the experience. Participants used concise and directive verbal exchanges to coordinate actions. Group reflections frequently referenced the need to improve the clarity and timing of communication. Teams that communicated effectively (e.g., Pair 8, Group 2) also performed better. This echoes previous findings where players felt more connected when their work was tightly coupled and involved intense verbal exchange \cite{harris_asymmetry_2019, rogers_best-fit_2021}.

Shared View as coordination mode frequently co-occurred with real-time verbal exchange and can be interpreted as an adaptive coordination strategy, especially when it was not overused (e.g., Pair 1, 2, 3). However, extended reliance on Shared View appeared to signal moments where verbal description alone was insufficient. Pairs 4 and 6 shared screens prominently in Task 2 of \textit{Lost in Space}, which required accurate descriptions of constellations and spatial relationships. Preferences for Shared View may reflect individual and team-level factors such as verbal communication skills, familiarity with spatial description, or previous gaming experience. In general, Shared View can indicate the need for a fallback strategy when purely verbal coordination leads to overload.

\highlight{The communication patterns observed in our study, including rapid and shorthand “gamer communication,” raise questions about their transferability to broader contexts. While such styles enabled efficient coordination in our sample, they rely on shared conventions, prior experience, and a willingness to engage in fast-paced, informal dialogue. In organizational settings with more diverse teams or lower familiarity with interactive systems, these patterns may not emerge naturally. This suggests that additional scaffolding, such as communication prompts or facilitation, may be necessary to support verbal coordination beyond technically experienced groups.}

\subsubsection{Information asymmetry (F5) and spatial separation (F8)}
Information asymmetry was designed to enforce interdependence and collaborative problem-solving. As shown in the results, asymmetry compelled coordination in most cases. Players often had to describe visual clues, explain tools, or request actions from teammates and were aware of the importance of these actions. However, spatial separation also introduced challenges to awareness. 

In the second task in \textit{Lost in Space}, the team members had more flexibility to split their roles. Following Harris \cite{harris_asymmetry_2019}, we envisioned that this can lead to forms of "emergent cooperation", which occurred for most pairs. The exception is Pair 7, where one user played alone. This provided fertile ground for reflection in the group discussion. On a positive note, we see that the task created the context for teamwork conversations. However, information asymmetry was less supported for the triangulation task in \textit{Lost in Space} than for Task 1 in \textit{Lost in Space} and in \textit{The Journey}. 

\highlight{We used asymmetry to introduce desirable difficulty in the scenarios. Increased task complexity and time pressure in \textit{The Journey} supported tighter collaboration, but such conditions may exceed the comfort threshold of some participants, particularly those with less gaming experience. This reflects a well-known trade-off in game design: balancing challenge and skill. While increased challenge can make coordination necessary and visible, it must be balanced with accessibility to avoid disengagement or unequal participation. For real-world deployment, this requires careful calibration of task difficulty in relation to participants’ prior experience and expectations.}

\subsubsection{F7: Inventory and tool exchange}
The tool inventory and the possibility of exchanging tools were meant to facilitate role negotiation and joint action. As can be seen in the findings, the tools were usually split between the participants, signaling a clear division of tasks. However, the exchange of tools was not used to its full potential. Players usually remained in possession of the tools that they initially collected, and only infrequently changed ownership between them. In this sense, users did not actively avoid being locked in a single dynamic during gameplay \cite{harris_asymmetry_2019}. However, it was possible for one team member to effectively monopolize the tools. A potential safeguard is to limit the number of tools a player can hold to prevent dominance. 

\subsubsection{F9: Structured Reflection}
Through structured reflection, our goal was to support adaptive behavior. Our handouts were part of "outmersive game design" \cite{miller_design_2024}: they were conceived outside of the experience to create a distance when revisiting the team process. We found this to be effective in prompting participants to analyze team dynamics. The debriefs led to concrete suggestions to improve task division, communication strategies, and leadership roles. We also dealt with uncomfortable topics, such as dealing with failure, thus transferring from the game experience to the real world \cite{miller_design_2024}. Facilitators played a key role in creating a space for teams to process these unplanned findings \cite{kolbe_how_2016}. 

\subsection{RQ2: Development of KSAs through gameplay}
\highlight{While team development is a long-term process, our findings indicate that participants engaged in and practiced key teamwork skills within the structured workshop setting. Rather than evidencing sustained development, we focus on how these skills were enacted during the experience, based on the collected data.}

\subsubsection{Communication skills}
The design of the experiences deliberately relied on verbal coordination to promote open, task-focused communication (KSA7) and listening behavior (KSA8) \cite{stevens_knowledge_1994}. Information asymmetry and spatial separation forced players to describe their environments, interpret clues, and align their actions through speech. Typically, players initially struggled to orient themselves at the beginning of Scenario 1, but in both experiences they reported only average levels of awareness of each other’s actions. This indicates that the players did not fully know what the other players were doing. 

The questionnaire data showed moderate to strong agreement on communication quality, with participants reporting higher satisfaction and perceived contribution to The Journey, compared to Lost in Space. These improvements suggest that the progression between the two scenarios helped the players become more comfortable with verbal coordination. 

Group reflections revealed that players often adopted a "gamer communication" style: fast, efficient, and minimal. Although this supported real-time action, it occasionally limited deeper understanding, leading to misalignment or duplication. Detailed descriptions of actions and the environment were prevalent when players were unsure how to proceed. As tasks became more complex in \textit{The Journey}, players recognized the need for clearer communication strategies and verbal feedback loops (KSA7, KSA8, KSA21). 

The video data confirmed that tightly coupled segments often involved real-time verbal exchange. Pair 8 (Lost in Space), the fastest to complete the task, demonstrated sustained and effective verbal interaction without relying on shared screen views, indicating clear communication and active listening. Pair 7, in contrast, demonstrated the difficulty in successfully completing missions when communication is limited. In general, reduced awareness did enforce the need for verbal exchange. \highlight{However, if players are not used to verbal descriptions, it may lead to a feeling of disconnection. A future study with a larger sample can further address whether this is influenced by personal or contextual factors}. 

\subsubsection{Coordination and collective efficacy}
The game required coordination through joint planning, interdependent task flows and collective adaptation - supporting KSAs related to goal alignment (KSA13), role clarity (KSA14) and collective efficacy (KSA15–17, KSA19). Data from the questionnaire indicated a strong awareness of task interdependence. In The Journey, 93.3\% of the participants reported dividing tasks, an increase from 81\% in \textit{Lost in Space}, and rated effectiveness and contribution higher. These findings suggest improved coordination and growing collective confidence while progressing in the game, with respect to achieving shared goals (KSA19). Insights from group discussions suggest that this shift was driven by the increased challenge and interdependence of the second scenario, which appeared to be more engaging. In the first scenario, while we directly observed that all participants collaborated to solve tasks, with the exception of Pair 7, the perception of collaboration diverged. 31.25\% of the participants did not feel that they collaborated closely, a response echoed by some participants that described the collaboration demands as relatively straightforward. In effect, these participants found that the first scenario was relatively easy and did not perceive strong collaboration compared to other multiplayer games they had tried. We therefore interpret this divergence as reflecting differences in participant expectations and prior experience. However, it also points to the potential of adaptive gameplay to better address differences in knowledge and skills.

In terms of coordination strategies, group discussions revealed that some teams used fixed roles, while others adapted. Participants noted when roles were unbalanced and discussed ways to improve this, for example, by coming up with alternative strategies for task division. {We see these as forms of "emergent coordination" that were enabled by our experience design that allowed flexibility in task division \cite{harris_asymmetry_2019}. These reflections illustrate an emerging awareness of team capacity and mutual support (KSA16, KSA17).

The video data reinforced these insights. Successful groups (e.g., Pair 1, Group 2) showed frequent role switching, tool exchange, and verbal confirmation of goals, suggesting effective activity synchronization (KSA13) and role negotiation (KSA14) \cite{stevens_knowledge_1994, shuffler_theres_2011}. In contrast, Pair 7 demonstrated poor coordination, with one player dominating the task.

\subsubsection{Trust and cohesion}
Trust and cohesion were nurtured indirectly through mechanics that required players to rely on each other for access to tools, information, or task progression. The responses to the questionnaire suggest moderate levels of trust-related behavior. The satisfaction reported by the participants, especially in \textit{The Journey}, points to growing confidence in the contributions of the teammates. We saw cohesion emerge in group discussions, as players began to reflect on their sense of being a team \cite{sottilare_designing_2018}. Trust was also tested when responsibilities were uneven, indicating that the design of the experience can support or hinder shared ownership of the task (KSA21).

The video data highlighted how trust was enacted in both tight and loose coordination modes. For example, participants often relied on their partner to perform unseen actions or describe visual elements. These behaviors imply confidence in the accuracy and intent of the other (KSA21). Teams that frequently exchanged information when they were further apart demonstrated implicit trust, while those who preferred tighter modes worked in close coordination and decision making.

\subsubsection{Reflexivity}
The ability to evaluate and adapt strategies was supported both within the experience and through structured debriefs. Group reflections revealed clear instances of team-level reflexivity. The participants reviewed their initial coordination strategies and proposed more efficient future approaches. Following previous research, we saw that structured reflection handouts led players to articulate successes, breakdowns, and adjustments, targeting KSA27 and KSA28 \cite{gurtner_getting_2007}. We observed that the time allocated for reflection within groups is important for teams to review their actions and leads to better adaptation strategies \cite{konradt_reflexivity_2016, gurtner_getting_2007}.

\subsubsection{Breakdowns and design limits}
We observed one case of collaboration breakdown (Pair 7), where one participant solved the task individually and later acknowledged their competitive behavior during the group discussion. This breakdown surfaced individual coordination preferences that may remain hidden in more constrained tasks (e.g., Task 1). This case points to a design trade-off. Tasks with weaker asymmetry, such as Task 2, offer greater flexibility in how teams coordinate, but do not prevent individuals from overtaking the mission. However, this openness also limits the extent to which collaboration is structurally enforced. As a result, such tasks are more likely to reveal tendencies toward competitive or individual problem-solving.

Breakdowns can become productive learning moments only when participants are willing and supported to reflect on their own behavior. In our workshop setting, structured reflection provided the space for this process. This raises a question for future work: How could reflection be supported without a human facilitator, for example, through structured prompts, guided self-reflection, or AI-based coaching systems.

The responses of the questionnaire suggest that reflection played a role in increasing effectiveness between rounds. The players iterated on earlier strategies, showing qualitative and quantitative improvements in communication \cite{konradt_reflexivity_2016}. The video data showed behavioral evidence of reflexivity as participants adapted their coordination styles mid-game. Typically, sharing mental models is more common in transition phases \cite{burke_understanding_2006}, but our asymmetric design made this critical to solving the task, which led to more intensive practice of this behavior. For example, Pairs 4 and 6 increased their use of shared views during the more complex second half of \textit{Lost in Space}, suggesting that the players adjusted their coordination mode based on task demands (KSA28).

\subsection{Design insights and recommendations}
Based on our findings, we propose a set of recommendations for researchers, designers, and practitioners who develop immersive experiences that aim to train teamwork KSAs such as communication, coordination, trust, and reflexivity. Although our design recommendations are grounded in observed behaviors, they are most immediately relevant for training contexts involving students, early-career professionals, or apprentices, and should be adapted when applied to more diverse organizational settings. \highlight{Moreover, they are relevant for short interventions aimed at practicing teamwork behaviors and skills.} In the following, we identify the \textbf{element of the game design} responsible for learning certain skills of team training \textbf{skills} and formulate our recommendations on how to implement it based on our findings.

\subsubsection{Design for purposeful verbal communication}
One of the core mechanics used in the experience was the combination of spatial separation (F4), information asymmetry (F5), and tool dependencies (F6). This design specifically aimed to develop \textit{communication skills} such as clarity of communication (KSA7), active listening (KSA8), and task coordination (KSA13). Tasks were intentionally structured so that players had access to non-overlapping information, requiring them to coordinate verbally in order to progress. By limiting visual access to shared resources and distributing clues among participants, the design encouraged players to develop a shared understanding through dialogue. Although other asymmetric games did not specifically tackle team development \cite{harris_beam_2015, harris_asymmetry_2019}, we found that players progressively improved their use of verbal strategies. However, reliance on verbal exchange alone presented challenges: some participants experienced cognitive overload or struggled to describe visual information clearly, adding to previous concerns about tension and fatigue in asymmetric conditions \cite{goncalves_exploring_2021}. In such cases, teams often defaulted to shared view interactions, using the same tablet screen to re-establish common ground. These findings suggest that, while verbal coordination is effective, designers should consider providing fallback mechanisms for users who may find sustained verbal communication overwhelming.

\subsubsection{Embed interdependence through asymmetric tools and roles}
The experience incorporated asymmetric design by assigning unique tools and information to each player, reinforced through inventory mechanics (F5) and the need for tool hand-offs. This approach supports \textit{coordination skills}, such as collective efficacy (KSA15), mutual support (KSA17), and balanced workload distribution (KSA14). Where the task allowed some flexibility, most teams cooperated successfully. However, there were also pitfalls: When one player held a dominant tool or role for too long, their teammate disengaged. To counteract this, designers can restrict tool access or functionality based on role or context. However, hard-coded role enforcement can inadvertently constrain organic team dynamics and learning opportunities. Others have found that players prefer to alter the degree of interdependence from their partners \cite{harris_asymmetry_2019}. More effective are design solutions that embed interdependence intrinsically into the gameplay, prompting role sharing and mutual reliance without explicitly prescribing it.

\subsubsection{Use narrative and role metaphors to build team identity}
To foster a sense of shared purpose and cohesion, the experience was embedded within a narrative structure that cast the participants as members of the crew on a collective mission. This background story, combined with metaphorical roles, supported the development of team cohesion (KSA24) and reinforced a sense of shared identity and motivation (KSA19). Using metaphors that aligned with team functions, the experience provided a framework to understand and implement collaborative roles. Importantly, this structure also allowed participants to explore roles, potentially gaining empathy for other perspectives. However, designers should be aware that not all metaphors resonate equally with all user groups \cite{gibson_metaphors_2001}.

\subsubsection{Incorporate structured reflection and progressive play}
To support deeper learning and team growth, the experience included structured reflection activities and replay opportunities. These elements were designed to foster team reflexivity (KSA27) and strategy adaptation (KSA28), both essential to improve collaborative performance \cite{konradt_reflexivity_2016, konradt_effects_2015}. The handouts and guided instructions provided outside the immersive experience \cite{miller_design_2024} encouraged the participants to evaluate their coordination, communication, and task division, while the progression of the experience allowed the teams to iteratively apply and refine their approaches. However, reflection activities are only effective if participants are actively involved. The reflection process can focus on the experience itself, through in-game prompts \cite{kleinman_locked_2025}, but when reflections are meant to extrapolate to other daily experiences such as teamwork, handouts, or facilitators \cite{miller_design_2024} may ensure a critical examination of behaviors and transfer.

\section{Limitations and Future Work}
Although our findings provide valuable information for the design of VR experiences based on tablets for team development, several limitations should be considered when interpreting the results and applying them in other contexts.

First, our work is grounded in a comprehensive framework of teamwork KSAs (Table \ref{tab:ksa_categories}). However, the study operationalizes and examines only a subset of these constructs. We selected KSAs that emerged most prominently in the expert interviews and could be meaningfully enacted within short, co-located, asymmetric VR scenarios. Consequently, the findings primarily inform the design and evaluation of immersive experiences that target foundational teamwork skills (e.g. KSA7, KSA8). These skills constitute the core building blocks of effective teamwork.

Other KSAs (e.g., conflict resolution, negotiation, or trust) were not explicitly addressed as they typically require specific design mechanics, such as intentional breakdowns, competing goals, or adversarial dynamics. For example, to address trust, we envision a scenario inspired by prisoner’s dilemma–style games, in which players can only succeed through cooperation, while key aspects of the rules remain hidden and require reliance on teammates. To target conflict resolution, future scenarios could deliberately introduce competing objectives or resource scarcity, prompting players to manage opposing interests. Expanding on our current work with such additional scenarios would enable the design of long-term interventions consisting of progressive scenarios that address the full set of 28 KSAs.

Second, due to a small and homogeneous sample of university students with relatively high familiarity with games and digital tools, the results may not generalize to broader workplace demographics, especially non-technical or older populations. Consequently, the proposed design recommendations should be interpreted as directly applicable to younger participants, such as students, early-career professionals, or apprentices. Many organizations invest in the upskilling of such young employees and trainees, for whom immersive and game-based learning approaches may be particularly effective \cite{noauthor_future_2025}.

Previous work suggests that for qualitative research, a sample size of 12 participants suffices when researching interactive systems \cite{caine_local_2016}, while other estimates suggest that nine to ten participants are enough for theme saturation \cite{wutich_sample_2024}. As we triangulated data from three sources, we believe that the study presents consolidated findings for the research context of team development for a young and digitally educated group. 

Third, the workshop lasted three hours, including play sessions and reflection. Although this duration allowed us to observe immediate collaboration behaviors, it did not capture how teamwork skills evolve through repeated exposure or longitudinal use. Consequently, we do not claim sustained team development effects. Rather, our findings demonstrate how short and well-structured immersive experiences can create conditions in which foundational teamwork KSAs are actively enacted, rehearsed, and reflected upon. Previous work  shows that even brief, well-designed interventions can produce meaningful improvements in teamwork processes (e.g. \cite{salas_does_2008}). Such interventions can initiate learning by making coordination demands visible and supporting reflection on team interaction patterns. Hence, our contribution lies in showing how immersive design can prompt and structure behaviors that are foundational to the development of teamwork-related skills.

Our study did not include a control group or a comparison to non-immersive conditions as a deliberate methodological choice aligned with the exploratory and qualitative nature of the work. Our goal was to examine how specific immersive design elements, such as asymmetric information, spatial separation, and interdependent tools, shape the enactment of teamwork KSAs. We prioritized in-depth observation and analysis to generate design knowledge that can inform future controlled studies. Our findings should be interpreted as formative insights into collaboration dynamics within immersive asymmetric systems.

\section{Conclusion}
\highlight{We presented the design process and evaluation of a tablet-based VR experience that supports the enactment of teamwork-related KSAs. Based on interviews with HR professionals and informed by established KSA frameworks, our design incorporates asymmetry, verbal coordination, and interdependent gameplay as key mechanisms to foster collaboration.}

\highlight{Our findings illustrate how a narrative-driven VR environment can structure the practice of teamwork skills in short interventions, and how participants enact these KSAs through coordinated actions and strategic dialogue.}

\highlight{By intentionally designing for interdependence and reflection, we contribute actionable insights for researchers and practitioners seeking to create conditions that support the practice of teamwork skills. As immersive technologies become more accessible in organizational settings, future work is needed to examine how such designs translate to different contexts, populations, and longer-term interventions.}

\bibliographystyle{ACM-Reference-Format}
\bibliography{Bibliography}

\end{document}